\documentclass{aa}
\usepackage{graphicx}
\usepackage[varg]{txfonts}
\usepackage{enumitem}
\usepackage{amsmath,amssymb}
\usepackage{booktabs}
\usepackage{xspace}
\usepackage{textcomp}
\usepackage{xcolor}	
\usepackage[euler]{textgreek}
\usepackage{dirtytalk}
\usepackage{tabularx}
\usepackage{caption}
\usepackage{subcaption}
\usepackage{placeins}
\usepackage{hyperref}

\graphicspath{{./figures}}

\newcommand{\GIZMO}{{\sc Gizmo}}

\newcommand{\SIMBA}{{\sc Simba}}
\newcommand{\Despotic}{{\sc Despotic}}
\newcommand{\SLICK}{{\sc Slick}}
\newcommand{\CAESAR}{{\sc caesar}}
\newcommand{\Zsun}{{\rm Z}_{\odot}}
\newcommand{\Msun}{{\rm M}_{\odot}}
\newcommand{\alphaCO}{\alpha_{\rm CO}}
\newcommand{\HI}{\ion{H}{I}}
\newcommand{\Mstar}{M_*} 

\begin{document}

   \title{Predicting the resolved CO emission of $z=1-3$ star-forming galaxies}
   
   \titlerunning{Predicting the resolved CO emission of $z=1-3$ star-forming galaxies}

   \author{Ravishankar Anirudh\inst{1,2}, Melanie Kaasinen\inst{3}, Gerg\"o Popping\inst{3}, Desika Narayanan\inst{4,5}, Karolina Garcia\inst{4,6}, Dariannette Valentin-Martinez\inst{4} }
    
    \authorrunning{Anirudh, et al.}    

   \institute{Max-Planck-Institut f{\"u}r Astronomie, K{\"o}nigstuhl 17, D-69117 Heidelberg, Germany\\
              \email{ravishankar@mpia.de}
            \and
          Indian Institute of Science Education and Research (IISER) Tirupati, Rami Reddy Nagar, Karakambadi Road, Mangalam (P.O.), Tirupati 517 507, India
          \and
          European Southern Observatory, Karl-Schwarzschild-Str. 2, D-85748, Garching, Germany
          \and
          Department of Astronomy, University of Florida, 309 Bryant Space Science Center, Gainesville, FL 32611, USA  
          \and 
         Cosmic Dawn Center at the Niels Bohr Institute, University of Copenhagen\
 and DTU-Space, Technical University of Denmark
          \and
          Center for AstroPhysical Surveys, National Center for Supercomputing Applications, University of Illinois Urbana-Champaign, Urbana, IL 61801, USA
    }

        \date{Received 18 November 2024; accepted 14 May 2025}

      \abstract
       {Resolved observations of the CO emission from z=1-3 star-forming galaxies are becoming increasingly common, with new high-resolution surveys on the horizon.
       }
       {We aim to inform the interpretation of this resolved CO emission by creating synthetic observations and testing to what extent routinely observed CO transitions can be used to trace H$_2$ across galaxy disks.
       }
       {To this end, we extract $z=1-3$ massive star-forming galaxies (on and above the main sequence) from the \SIMBA\ cosmological simulation and predict their spatially resolved CO(1--0)-to-CO(5--4) emission using the \SLICK\ pipeline, which combines sub-resolution modeling of the cloud population with the \Despotic\ spectral line calculation code.}
       {We find that the CO(1--0)-to-H$_2$ ratio ($\alphaCO$) varies significantly within these galaxy disks---from values of $\sim$1--5 $\mathrm{M_\odot}$ (K km s$^{-1}$ pc$^2$)$^{-1}$ in the central 1--3 kpc of the most massive galaxies to $>100$ $\mathrm{M_\odot}$ (K km s$^{-1}$ pc$^2$)$^{-1}$ at $\sim$ 15 kpc. Thus, the use of a single $\alphaCO$ to derive the H$_2$ surface density leads to severe underestimates of the H$_2$ contribution in its outskirts. As expected, higher-$J$ CO transitions trace molecular gas in the centers at higher densities, whereas CO(1--0) better traces the more diffuse, extended molecular gas. We see significant variations in the CO excitation, with CO(3--2)/CO(1--0) line luminosity ratios of the most massive galaxies at $z\sim$2 declining from $\sim$ 0.9 in the galaxy centers to $\sim$ 0.1 in the outskirts. On average, line ratios increase substantially toward higher redshifts and lower galaxy stellar masses.}
       {We predict that tracing molecular gas with CO beyond 3-5 kpc of cosmic noon galaxies will be challenging with current facilities due to the drastic increase in $\alphaCO$. On average, the half-light radii of all CO transitions up to CO(5--4) are consistent with each other, but are $\sim27\%$ smaller than the radii enclosing half the total H$_2$ mass. The predicted line ratios for the central few kiloparsecs of massive galaxies reach supra-thermal values in warm ($\sim30-100$\,K), dense ($\mathrm{>100\,cm^{-3}}$) gas. The increased fraction of dense gas in galaxy centers and toward higher redshifts gives rise to CO excitation gradients.}


     \maketitle

    \section{Introduction}
        \label{sec:intro}
    
        Resolving the molecular gas in $z$=1--3 (cosmic noon) galaxies is critical to studying the sites and conditions for star formation during its peak epoch. Since the dominant component of molecular gas, H$_2$, barely emits at the low temperatures typical of this star-forming gas, observational studies---both near and far---have come to rely on rotational transitions of the carbon monoxide  molecule (CO), which are typically the brightest lines from cold gas accessible at cosmic noon. Yet, it is still unclear exactly how this CO line emission scales with the underlying H$_2$ distribution beyond $z>0$.

        For local galaxies, H$_2$ is commonly traced via the ground transition, CO(1--0), but at cosmic noon, CO(1--0) has been resolved for only a handful of galaxies with the Very Large Array 
        \citep{Riechers_2011, Sharon_2013,Aravena_2014, Spilker_2015, Stanley_2023}, with higher rotational quantum number ($J>1$) transitions now more commonly resolved with submillimeter interferometers \citep[e.g.,][]{Tadaki_2017,Calistro_2018,Puglisi_2019,Kaasinen_2020,Lelli_2023,Rizzo_2023}. Not only is the targeted transition changing, but so too are the types of galaxies being resolved. Most of the early, resolved studies at $z>1$ (with the VLA and IRAM Plateau de Bure Interferometer) targeted the brightest, dust-rich and/or lensed systems. But now, the increase in resolution and sensitivity afforded by the Atacama Large Millimetre/submillimetre Array (ALMA) is enabling resolved observations of molecular gas in more typical, yet still massive ($\Mstar\gtrsim 10^{10.5}\,\Msun$), main-sequence galaxies \citep[e.g.,][]{Kaasinen_2020,Ikeda_2022, Puglisi_2021,Tadaki_2023}.
        
        Using resolved ($\sim$ few kiloparsecs) observations of molecular gas at $z$=1--3, various studies have quantified the ``size'' of the CO emission compared to other galaxy components, usually parameterized in terms of the half-light radius \citep[e.g.,][]{Chen_2017, Calistro_2018, Puglisi_2019, Kaasinen_2020}. For example, multiple studies have found the CO half-light radius to be smaller than that of the 1.6\,\textmu m stellar emission, yet larger than that of the 850\,\textmu m--1.3\,mm dust-continuum emission \citep{Chen_2017, Tadaki_2017, Puglisi_2019, Calistro_2018, Kaasinen_2020,Ikeda_2022,Rybak_2024}. This mismatch in sizes is best interpreted through detailed physical models. Simulation-based, synthetic observations have already shown that the compact dust versus stellar emission can be attributed to a combination of heating and extinction effects \citep[e.g.,][]{Cochrane_2019, Popping_2022}. However, with no systematic predictions of the extent of CO emission, it is still unclear what physical properties the CO sizes are tracing, be it the H$_2$ density distribution, interstellar radiation field, gas-phase metallicity, and/or dust column density.
        
        Resolved CO observations have been used to trace radial variations in the molecular gas surface density \citep[e.g.][]{Sharon_2019} and the disk dynamics \citep[e.g.,][]{Lelli_2023, Rizzo_2023}, assuming that the CO surface brightness uniformly follows the mass distribution of H$_2$ across these galaxies. Yet, this is not the case for many local galaxies \citep{Sandstrom_2013,Hunt_2023,Chiang_2024}, and the extent to which this assumption may bias results at $z$=1--3 is uncertain. Moreover, the validity of this assumption for different rotational transitions remains unclear. For $z$=1--3 galaxies, various transitions have been observed at resolutions equivalent to 3--6\,kpc scale---especially CO(2--1), CO(3--2), and CO(5--4), which are commonly targeted with ALMA. However, there are still few sources with resolved observations of multiple CO transitions, enabling a comparison of the extent of these tracers. The two $z=3.4$ and $z=2.78$ galaxies studied in \cite{Riechers_2011} and \cite{Spilker_2015}, respectively, show tentative evidence of the CO(1--0) emission being more extended than the CO(3--2), (4--3) and (6--5) emission, whereas for the $z\sim2.2$ main-sequence galaxy in \cite{Bolatto_2015}, the CO(1--0) and CO(3--2) half-light radii appeared consistent. To understand how representative these results are, and what they imply about the underlying physical properties of the molecular gas, the community needs a set of reliable synthetic observations of the resolved CO emission in cosmic noon galaxies.
        
        Predicting the resolved emission from CO in $z$=1--3 galaxies has been challenging. Current cosmological simulations, and even isolated simulations of massive galaxies have lacked the resolution and physics needed to form molecular gas.  Isolated galaxy/zoom-in simulations have only recently begun reaching the resolution needed to directly capture the formation of CO and H$_2$ \citep{Richings_2016,Tress_2020,Richings_2022,Thompson_2024}. However, such approaches are only now being applied to cosmological simulations. Thus, most studies based on cosmological simulations have resorted to predicting CO emission through a combination of sub-resolution models (describing the distribution and properties of molecular clouds and the heating mechanisms) and radiative transfer models  (see e.g. \citet{Popping_2019} for an overview). Whereas the sub-resolution models can be adopted either on-the-fly or in post-processing, radiative transfer models are mainly applied in post-processing using spectral synthesis or photodissociation region (PDR) codes. 
         
        Sub-resolution approaches have already been used to predict the integrated CO line luminosities of galaxies in cosmological contexts. Examples include studies of $z=0$ galaxies with major cosmological simulations (e.g. with FIRE-2, \citealt{Keating_2020}; with IllustrisTNG, \citealt{Inoue_2020}; and with \SIMBA, \citealt{Olsen_2021}), studies of $z\ga6$ galaxies with cosmological zoom-in simulations \citep[e.g.][]{Munoz_2013,Vallini_2018,Schimek_2024}, and the coupling of semi-analytic models with sub-resolution molecular line radiative transfer modeling \citep[e.g.,][]{Lagos_2012,Popping_2014,Popping_2016,Popping_2019}. Moreover, numerous studies have predicted the integrated and resolved CO line emission from submillimeter galaxies using non-cosmological disk setups \citep[see][]{Narayanan_2009, Narayanan_2011, Narayanan_2012, Narayanan_2014,Bournaud_2015}. Recent advances have even enabled simulations to be combined with non-equilibrium chemistry solvers to predict the CO emission of local GMCs \citep[e.g.][]{Thompson_2024}.
        
        Yet, simulation-based predictions of the resolved CO emission in cosmic noon galaxies remain scarce, with only a few examples, such as \cite{Olsen_2016} and \cite{Greve_2008}. Although these studies provide useful comparisons, they are limited to small samples; \cite{Olsen_2016} simulated three main-sequence disk galaxies at $z \sim 2$, and \cite{Greve_2008} focused on a single Lyman-break galaxy at $z = 3$. Thus, there are still insufficient statistical estimates for z=1--3 galaxies predicting how the extent of CO emission compares to the H$_2$ mass distribution, and how the CO line excitation varies spatially with the star-formation activity and dust/gas properties.
        
        In this work, we provide statistical predictions for the surface brightness and total extent of CO emission in $z$=1--3 star-forming galaxies. To achieve this, we take hundreds of $z$=1--3 star-forming galaxies from the \SIMBA\ cosmological hydrodynamical simulations \citep{Simba} and apply the Scalable Line Intensity Computation Kit \citep[\SLICK,][]{Garcia_2024}, which combines the sub-resolution molecular cloud modeling of \cite{Popping_2019} and \citet{Narayanan_2017}, the chemical network of \cite{Gong_2017}, and the \Despotic\ line emission software \citep{Krumholz_2014}. Using this pipeline, \citet{Garcia_2024} successfully reproduced galaxy-scale observational constraints for the relationship between $\alphaCO$ and gas-phase metallicity at $z = 0$. We extend this work to create resolved, velocity-integrated maps of the CO(1--0)$\to$CO(5--4) emission of these simulated galaxies, using these synthetic data to test: 
        1) how the CO half-light radii relate to the H$_2$ distribution, 
        2) how well the CO surface brightness follows the H$_2$ surface density as a function of galactocentric radius, and 
        3) to what extent the CO excitation changes with galactocentric radius.  

        This paper is structured as follows. In Sec.~\ref{sec:method} we briefly describe the hydrodynamical simulations, selection of the simulated galaxies, and post-processing to produce line intensity maps via \SLICK. We present our results, including a comparison of half-light radii, surface brightness profiles, radial variations in the CO-to-H$_2$ conversion factor and its dependence on physical conditions in Sec.~\ref{sec:results}. We place our results in the context of observational studies, discussing the implications in Sec.~\ref{sec:discussion}, and summarize our findings in Sec.~\ref{sec:summary}. Throughout this paper, we adopt a flat $\Lambda$CDM cosmology with $\Omega_m = 0.3$, $\Omega_\Lambda = 0.7$, $\Omega_b = 0.048$, 
        $h = H_0/(100~\mathrm{km~s^{-1}~Mpc^{-1}}) = 0.68$, $\sigma_8 = 0.82$, and $n_{\rm s} = 0.97$.

\section{Methods}
    \label{sec:method}

    \begin{figure*}[ht!]
        \centering
        \includegraphics[width=0.49\textwidth]{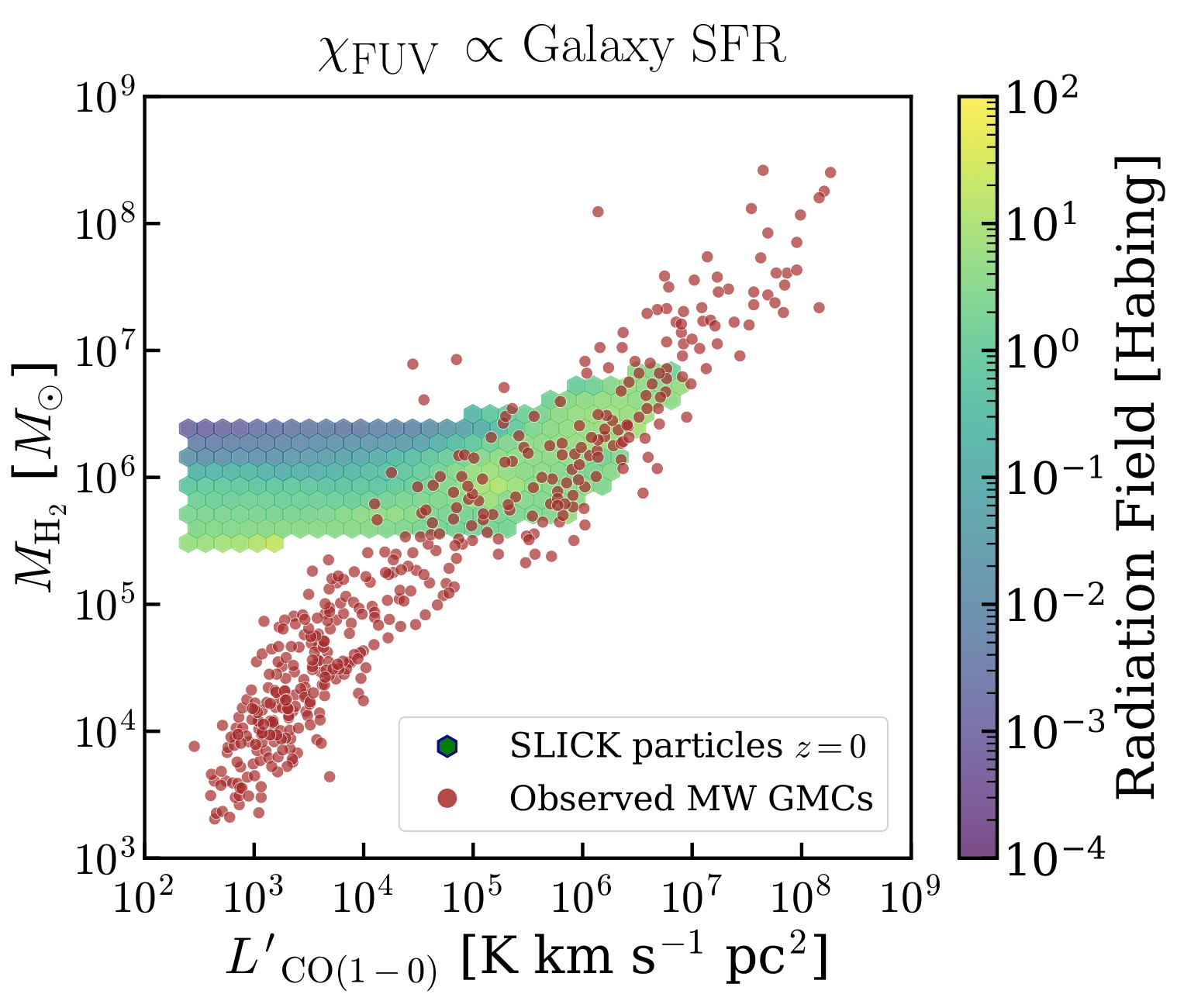}
        ~
        \includegraphics[width=0.49\textwidth]{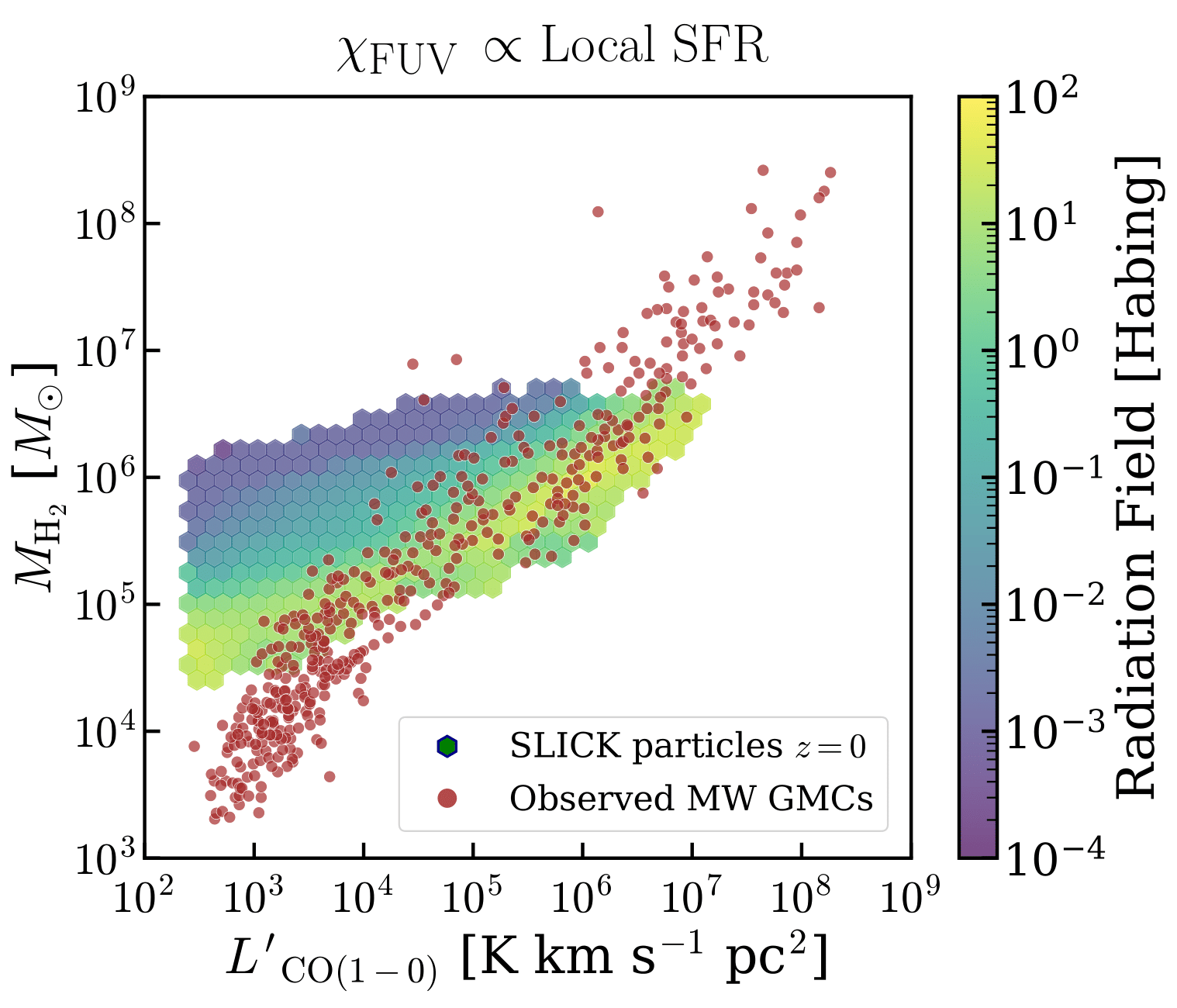}
        \caption{Different treatments of the interstellar radiation field, as shown through the correlation between the clouds' H$_2$ mass vs. CO(1--0) luminosity: the left-hand panel shows a non-localized implementation (i.e., same ISRF for all clouds in a galaxy) whereas the right-hand panel shows the localized treatment (considering the 64 nearest-neighboring clouds). \SLICK's clouds are indicated by the hexagonal bins, where the colormap indicates the strength of the radiation field impinging on the clouds. The localized ISRF approach better matches observed $z=0$ extragalactic GMCs \citep[red filled circles;][]{Bolatto_2008,Pineda_2009,meyer_resolved_2011, Wong_2011,Rebolledo_2012,meyer_resolved_2013}.
        \label{fig:newslickCOh2}}
    \end{figure*}

    \subsection{The \SIMBA\ cosmological simulation}
        
        To produce synthetic observations of a sizeable and representative $z$=1--3 galaxy sample, we use the \SIMBA\ cosmological galaxy formation simulations, run with \GIZMO's meshless, finite-mass hydrodynamics solver \citep{Hopkins_2015}. As detailed in \cite{Simba}, \SIMBA\ is based on a $\Lambda$-CDM cosmology in accordance with the results of \cite{Planck_collab} and includes several sub-resolution methods to trace the formation and evolution of stars, the chemical evolution of the interstellar medium (where the heating and cooling processes are calculated via the {\sc Grackle}-3.1 cooling library, \citealt{Smith_2017_Grackle}), and the formation, growth and feedback of black holes. As part of this, \SIMBA\ also tracks the abundance of 11 chemical species. Of particular relevance to our work are the treatment of the star formation rate (SFR), dust, and H$_2$. \SIMBA\ determines a simulated particle's SFR from the H$_2$ density, which is computed following the sub-resolution model of \cite{Krumholz&Gnedin2011}, using its metallicity and local column density. \SIMBA\ also accounts for on-the-fly formation, growth, and destruction of dust (in contrast to several other current cosmological simulations), improving the reliability of our UV attenuation models. In this way, \SIMBA\ successfully reproduces the sizes and H$_2$ mass fractions of star-forming galaxies, as well as various empirical relationships, including the stellar mass--SFR relation (i.e., the star-forming main sequence), and the mass-metallicity relation \citep{Simba}. By modeling dust on-the-fly, \SIMBA\ also predicts the evolution of the galaxy dust-to-gas ratio (DGR) and dust-to-metal ratio (DTM), providing a strong match to observations at $z\sim0$ \citep{QiLi_2019}, as well as $z\sim 6$ constraints \citep{Lower_2023,Lower_2024}.

        Available as snapshots and halo/galaxy catalogs, the \SIMBA\ simulation data\footnote{\url{https://simba.roe.ac.uk}} covers 151 redshifts from $\mathrm{z=20}$ to $\mathrm{z=0}$, for three different volumes. These volumes are defined by their box sizes of 25, 50, and 100 cMpc/h. In this work, we use the highest-resolution run, \SIMBA25, which consists of $2\times512^3$ particles with a gas particle mass resolution of $\mathrm{2.85\times10^{5}~\Msun}$ each. We obtain particle information (such as the gas mass and metallicity) directly from the simulation, assigning these to individual galaxies via the 6-D Friends-Of-Friends algorithm, \CAESAR\ \citep{Thompson_2015_caesar}. In this way, we obtain galaxy properties such as the center-of-mass (COM), gas and stellar masses, gas-phase metallicities, star formation rates (SFRs), and various radii enclosing 20, 50, and 80\% of the total mass of each particle type.        
    
    \subsection{The \SLICK\ modeling pipeline}
        \label{subsec:slick}
    
         To model the CO line emission of the simulated galaxies, we used the \SLICK\ framework, presented in \cite{Garcia_2024}. The details can be found in that paper. However, we summarize the basic workings of \SLICK\ below.

         \subsubsection{Inferring gas cloud sizes and densities}

        Cosmological simulations such as \SIMBA\ typically lack the internal density contrasts of molecular clouds that give rise to high-excitation CO emission. Therefore, extracting information about cold gas requires sub-resolution modeling. To infer the properties of clouds (or simply gas particles), \SLICK\ closely follows the approach presented in \cite{Popping_2019}. For each simulation snapshot (at redshift, $z$), the gas particles are extracted along with the simulation-derived particle mass, $M_\mathrm{C}$, number density, $n_\mathrm{C}$, and temperature, $T_\mathrm{C}$. The latter two quantities are merely used as a first guess to determine the cloud properties, and are recalculated later by \Despotic, to capture the density and temperature variations within each cloud. $M_\mathrm{C}$ of each particle is derived in \SIMBA\ by evaluating the smoothed density of the nearest 64 neighbors at the particle coordinates, each weighted by a cubic spline kernel as described in \cite{Hopkins_2015}.
             
             Using these simulation-based quantities, the external pressure was derived assuming an equation of state,
             \begin{align}
                P_\mathrm{ext} = n_\mathrm{C} k_\mathrm{B} T_\mathrm{C} \, ,
             \end{align}
             where $k_B$ is the Boltzmann constant. These pressure values, along with the cloud masses, are used to determine the cloud radii following \cite{Narayanan_2017} and \cite{Swinbank_2011}, via
             \begin{align}
                 \dfrac{R_\mathrm{C}}{\mathrm{pc}} = \left( \dfrac{P_\mathrm{ext}/k_\mathrm{B}}{10^4~\mathrm{cm^{-3}\,K}} \right)^{-1/4} \left(\dfrac{M_\mathrm{C}}{290~ \Msun}\right).
             \end{align}
             Each cloud is then assigned a power-law density distribution, divided into 16 concentric zones, with a hydrogen number density, $n_H$, given by,
             \begin{align}
                 n_H(R) = n_0 \left( \dfrac{R_C}{R} \right)^{2} \, .
             \end{align}
             where the innermost number density is given by,
             
             \begin{align}
                n_0 = \dfrac{M_\mathrm{C}}{4 \pi R_\mathrm{C}^3} \, .
             \end{align}

             Apart from the sub-resolution pressurization, the cloud is also assumed to be virialized. The total velocity dispersion $\sigma_\mathrm{tot}$, which is proportional to $M_{\rm C}/R_{\rm C}$ multiplied by the virial parameter, is computed assuming $\alpha_\mathrm{vir}=1$. Therefore, $\sigma_{\rm tot}$ only depends on the mass and radius of the cloud. 

         \subsubsection{Chemical network}
            \label{sec:chem}
             To compute the hydrogen and carbon chemistry of each zone, \SLICK\ uses the chemical network of \cite{Gong_2017} (which is already implemented in \Despotic). This network includes 18 chemical species: H, H$_2$, H$^+$, H$^{+2}$, H$^{+3}$, He, He$^+$, O, O$^+$, C, C$^+$, CO, HCO$^+$, Si, Si$^+$, e, CHx, and OHx. The C, O, and Si abundances are taken to be proportional to the gas-phase metallicity relative to the Solar Neighborhood, whereas the abundance of He is fixed to 0.1 per H nucleon. The dust abundance, dust cross section to the radiation fields, and the dust-gas coupling coefficient for energy exchange are scaled with the gas-phase metallicity and dust-to-metal ratio (DTM).

             \Despotic\ calculates the equilibrium chemical abundances of the 16 zones of each cloud by using the reaction rates of all species. The rate coefficients for the reactions are functions of gas temperature ($\mathrm{T_g}$), visual extinction ($\mathrm{A_V}$), and H nuclei density, and are are therefore iteratively calculated with $\mathrm{T_g}$. The tolerance for abundance convergence is set by the user, and we choose a value of $\mathrm{10^{-3}}$. Using the $\mathrm{H_2}$ abundance of all zones, we calculated the mass fraction of molecular hydrogen ($f\mathrm{H_2}$) for each cloud by using its particle mass $M_{\rm C}$. Hereafter, we refer to the molecular gas mass and $\mathrm{H_2}$ mass interchangeably, without accounting for the contribution of $0.4\times M_{\rm C}$ by helium to the molecular gas mass.

         \subsubsection{Radiation fields}
            \label{sub:radfields}
         
             To determine the line intensities, we first characterized several sources of radiation impinging on the molecular clouds: the interstellar radiation field (ISRF), cosmic ray (CR) field, and cosmic microwave background (CMB). We do not include the impact of X-rays in this work (deferring it instead to a follow-up study). Since X-rays dominate the excitation of CO at high-$J$ \citep{Meijerink_2007, vdW_2010,Rosenberg_2015,Kamenetzky_2017}, we restrict our study to $J_\mathrm{upp}\leq5$, and discuss comparisons to modeling efforts including AGN in Sec.~\ref{sec:discussion}.
             
             We account for the effects of the CMB via its temperature, which increases with redshift according to 
             \begin{align}
                T_\mathrm{CMB} = T_\mathrm{CMB}(z=0)~(1+z) = 2.73~(1+z) \, .
             \end{align}
             Whereas the CMB temperature does not change locally, the strength of both the ISRF and CR field do. These effects need to be taken into account to accurately predict the resolved emission from CO. 
             
             In \SLICK, the ISRF strength is parameterized via the UV field strength, $\chi$.  Early papers coupling {\sc despotic} to galaxy simulations \citep[e.g.][]{Narayanan_2014} determined the UV field strength from the global galaxy SFR.  This said, in {\sc SLICK}, we employ a localized approach, scaling the ISRF according to the local SFR density of each cloud, $\Sigma_\mathrm{SFR,C}$ \citep[following][]{Lagos_2012}. A similar approach was also used by \citet{Olsen_2016}. To achieve this, we adopt a nearest-neighbor approach, identifying the 64 nearest clouds/particles via SciPy's K-Dimensional Tree \citep{SciPy} module, using the particle coordinates from each \SIMBA\ snapshot. We tested this choice against neighbors selected according to different physical separations, but found that a number threshold was more appropriate than a distance threshold, as even distant particles created a sufficient UV flux to impact neighbors beyond these fiducial separations.
             
             We account for dust attenuation in our computation of $\chi$ by computing the dust mass density within the region covered by the 64 nearest neighbors, and converting it to a UV optical depth via
             \begin{align}
                \tau_{\rm UV} = \kappa_{\rm abs}\times \mathrm{R_{\sigma}\times \frac{3M_{dust}}{4\pi R_{\sigma}^3}},
                \label{eq:tau_UV}
            \end{align}
            where $\kappa_{\rm abs}$ is the mass attenuation coefficient/absorption cross section per g of dust in $\mathrm{cm^2\,g^{-1}}$ and $R_{\sigma}$ is the radius of the 64-particle sphere. We take $\kappa_{\rm abs} = 1.078 \times 10^5$\,cm$^{2}$\,g$^{-1}$, the median value of the mass attenuation coefficients within the range where CO and H$_2$ photo-ionization/dissociation/excitation usually occur, assuming a relative, Milky Way (MW) visibility of $R_v=3.1$.\footnote{\url{www.astro.princeton.edu/~draine/dust/extcurvs/kext_albedo_WD_MW_3.1_60_D03.all}} Using this optical depth, we computed the transmission probability of UV photons, $\beta_{\rm UV}$, as,
            \begin{align}
                \beta_{\rm UV} = \left(\frac{1-{\rm e}^{-\tau_{\rm UV}}}{\tau_{\rm UV}}\right)
                \label{eq:beta_UV}
            \end{align}
            following \cite{Lagos_2012}. We normalized this value by the Solar neighborhood value, determined using the MW gas surface density of 10 $\Msun\,\mathrm{pc}^{-2}$ \citep{Chang_2002} and a gas-to-dust ratio of 165, yielding $\beta_\mathrm{UV,MW}\simeq0.5$. Finally, the UV radiation field is determined as,
            \begin{equation}
                \chi = \dfrac{\Sigma_\mathrm{SFR,C}}{\Sigma_\mathrm{SFR,MW}} \times \dfrac{\beta_\mathrm{UV,C}}{\beta_\mathrm{UV,MW}} \, .
                \label{eq:attenuated_flux}
            \end{equation}
            where $\Sigma_\mathrm{SFR,MW}=790~\Msun\,\mathrm{Myr^{-1}\,kpc^{-2}}$ is the SFR surface density of the solar neighborhood \citep{Bonatto_2011}.   
            
             We determined the CR field strength, $\xi_\mathrm{CR,C}$, assuming that the sources setting the ISRF contribute proportionally. Since CRs are unattenuated (unlike the ISRF), we obtain
             \begin{align}
                 \xi_\mathrm{CR,C} = \dfrac{\Sigma_\mathrm{SFR,C}}{\Sigma_\mathrm{SFR,MW}} ~\zeta_{-16}~\xi_\mathrm{CR,MW}\, ,
             \end{align}
             assuming $\zeta_{-16}=0.1$ is the CR ionization rate and $\xi_\mathrm{CR,MW}=10^{-16}\mathrm{s}^{-1}$ is the CR field in the solar neighborhood, as in \cite{Narayanan_2014,Narayanan_2017}. In Fig.~\ref{fig:newslickCOh2}, we show the impact of our localized ISRF prescription, compared to the method using the global SFR used before. To highlight the differences, we show the correlation between the total H$_2$ mass of our galaxy sample, vs. the total CO(1--0) line luminosity. The new approach produces far more realistic cloud luminosities in the low mass regime than when using a globally averaged ISRF. The improvement is due to the $\chi$ in the global SFR approach being ineffective at destroying $\mathrm{H_2}$ (and CO), leading to a plateau of $\mathrm{M_{H_2}}$ in low/medium-density gas ($n_{\rm H}\lesssim10\,\mathrm{cm^{-3}}$) near star-forming regions. In contrast, the localized $\chi$ can reach much higher values (see colorbar in right panel), thereby resulting in better agreement with the $\mathrm{M_{H_2}-L'_{CO}}$ relation across all mass regimes.

         \subsubsection{Dust treatment} 

            The radiation calculations performed by \Despotic\ rely on several dust properties---including the dust abundance, dust cross-section to heating (by thermal radiation, 8--13.6 eV photons and ISRF photons), and the dust-gas coupling coefficient---which are all scaled with the dust-to-metal ratio (DMR). In \SLICK, we apply a variable DMR using the functional form derived by \cite{QiLi_2019} for \SIMBA\ galaxies. Their DMR scales with the dust-to-gas ratio (DGR), galaxy gas-phase metallicity, $Z_\mathrm{g}$, and the DMR of the Milky Way ($\mathrm{DMR_{MW}}=0.44$, \citealt{RemyRuyer_2014}) as,
            \begin{align}
                {\mathrm{DMR}} = \dfrac{\mathrm{DGR}}{Z_\mathrm{g}\cdot\mathrm{DMR_{MW}}}
            \end{align}
            where the DGR is given by,
            \begin{align}
                \mathrm{log\,DGR\,=\,2.445\,log\left(\frac{Z_\mathrm{g}}{\Zsun}\right)-2.029} \, .
            \end{align}

         \subsubsection{Computing the CO line luminosities with \Despotic}
         
            We make use of the python package {\sc despotic} \citep{Krumholz_2014} to model the chemistry and calculate the CO line luminosity of the individual clouds. {\sc despotic} simultaneously iterates over the thermal, chemical (Sec.~\ref{sec:chem}), and radiative equilibrium equations until convergence is reached in each of the concentric zones. Here, we summarize the relevant processes undertaken in calculating the CO line luminosities from each cloud. For a more detailed description of the set of equations that {\sc despotic} iteratively solves to calculate atomic and molecular species' luminosities, we refer the reader to \citet{Krumholz_2014} and \citet{Garcia_2024}.       
            
            The principal heating mechanisms considered in {\sc despotic} are the grain photoelectric effect, heating of dust by the interstellar radiation field, and cosmic ray heating of the gas.  The dominant cooling mechanisms are the molecular/atomic line emission and the thermal emission of dust grains.  Additionally,
            \Despotic\ includes cooling by Lyman $\alpha$ and Lyman $\beta$ lines and the two-photon continuum, using interpolated collisional excitation rate coefficients \citep{Osterbrock_2006}. Finally, the collisional exchange of energy between dust and gas is also taken into account, which becomes particularly relevant at densities of $n\ga 10^4$ cm$^{-3}$. 

            To solve for the statistical equilibrium within the level population of each atomic or molecular species, \Despotic\ applies the escape probability approximation. Density variations within a zone due to turbulence are accounted for by including a Mach-number-dependent clumping factor that represents the ratio between the mass-weighted and volume-weighted density of the gas. The clumping factor, $f_\mathrm{cl}$, depends on the cloud non-uniformity via the non-thermal velocity dispersion, $\sigma_\mathrm{non-thermal} = ( \sigma_\mathrm{tot}^2-\sigma_\mathrm{thermal}^2 )^{1/2}$. This value of $f_\mathrm{cl}$ is used as a multiplicative factor for calculating collisional rates. \Despotic\ accounts for the CMB both as a heating source (Sec.~\ref{sub:radfields}) as well as a background against which emission lines are observed.

    \subsection{Sample selection}

        To compare against recent observations, we predict the CO emission of massive, $z$=1--3 star-forming galaxies. We limited the sample to galaxies with a stellar mass above $10^9\,\Msun$ and a gas mass of $\geq 10^8\,\Msun$ (both of which are still far lower than most observed galaxies), and selected star-forming galaxies using the star-forming main sequence (MS) in \SIMBA\ determined by \cite{Akins_2022} for \SIMBA25. We only considered galaxies above their best-fit MS relation minus the scatter, taking 0.5\,dex below the relation as the minimum SFR cut-off per stellar mass. In this way, we selected main-sequence and starburst galaxies. We note that this sample selection naturally leads to a greater sample size at the end of cosmic noon, versus the beginning (that is, there are more galaxies meeting these criteria at $z$=0.8--1.2 than at $z$=2.5--3.5). We did not perform any cuts related to whether these galaxies host AGN. Our MS cut already excludes AGN hosts with strong jets, which are mostly quenched.
        
        We chose three bins equally spaced in redshift around $z$=1,2,3 and ran \SLICK\ for three snapshots in each bin. These are the $z$=0.8,1,1.2 ; $z$=1.8,2,2.2 ; $z$=2.5,3,3.5 snapshots respectively. The two higher redshift bins correspond to a total coverage of about 800 Myrs in cosmic time while the lowest redshift bin corresponds to about 2 Gyrs. Our initial analysis showed that many of the simulated galaxies were in the process of merging---complicating the analysis of their emission line profiles and sizes. To ensure that we derive accurate sizes and radial profiles for each galaxy, we removed galaxies that have more than one \CAESAR\ progenitor in the last 500 Myrs (i.e., which merged recently or are still merging). Although this cut is critical to our analysis, it has the effect of severely decreasing the sample size, particularly in the high-mass, high-redshift regime, leaving only one galaxy at $z$=2.5--3.5 with $\mathrm{M}_* > 10^{11} \Msun$. Given this limited number, we omit the results for this high-mass high-redshift bin, as it is unclear how representative this one galaxy is.

    \begin{figure*}[ht!]
        \centering
        \includegraphics[width=0.95\textwidth]{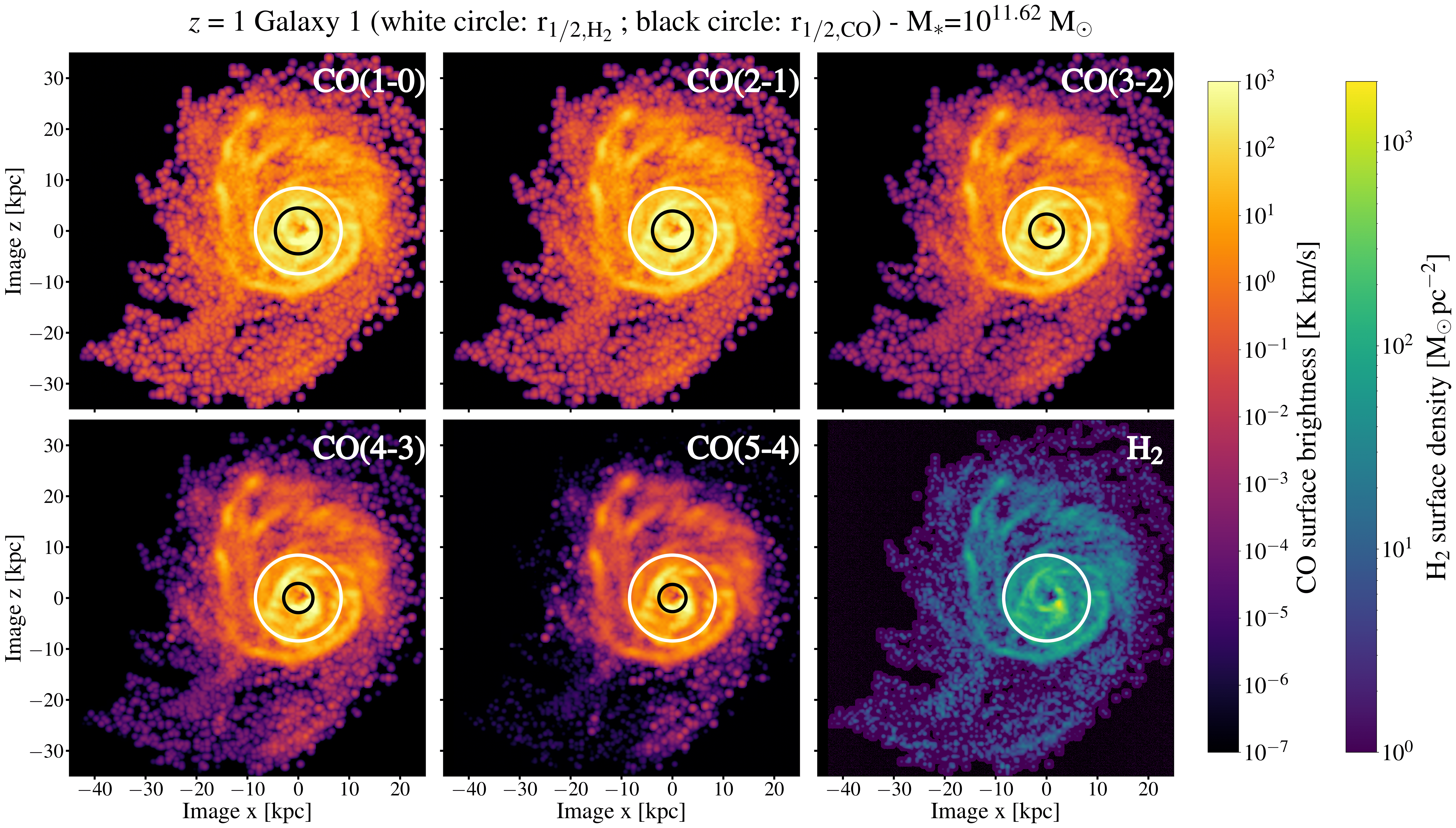}\\
        \includegraphics[width=0.95\textwidth]{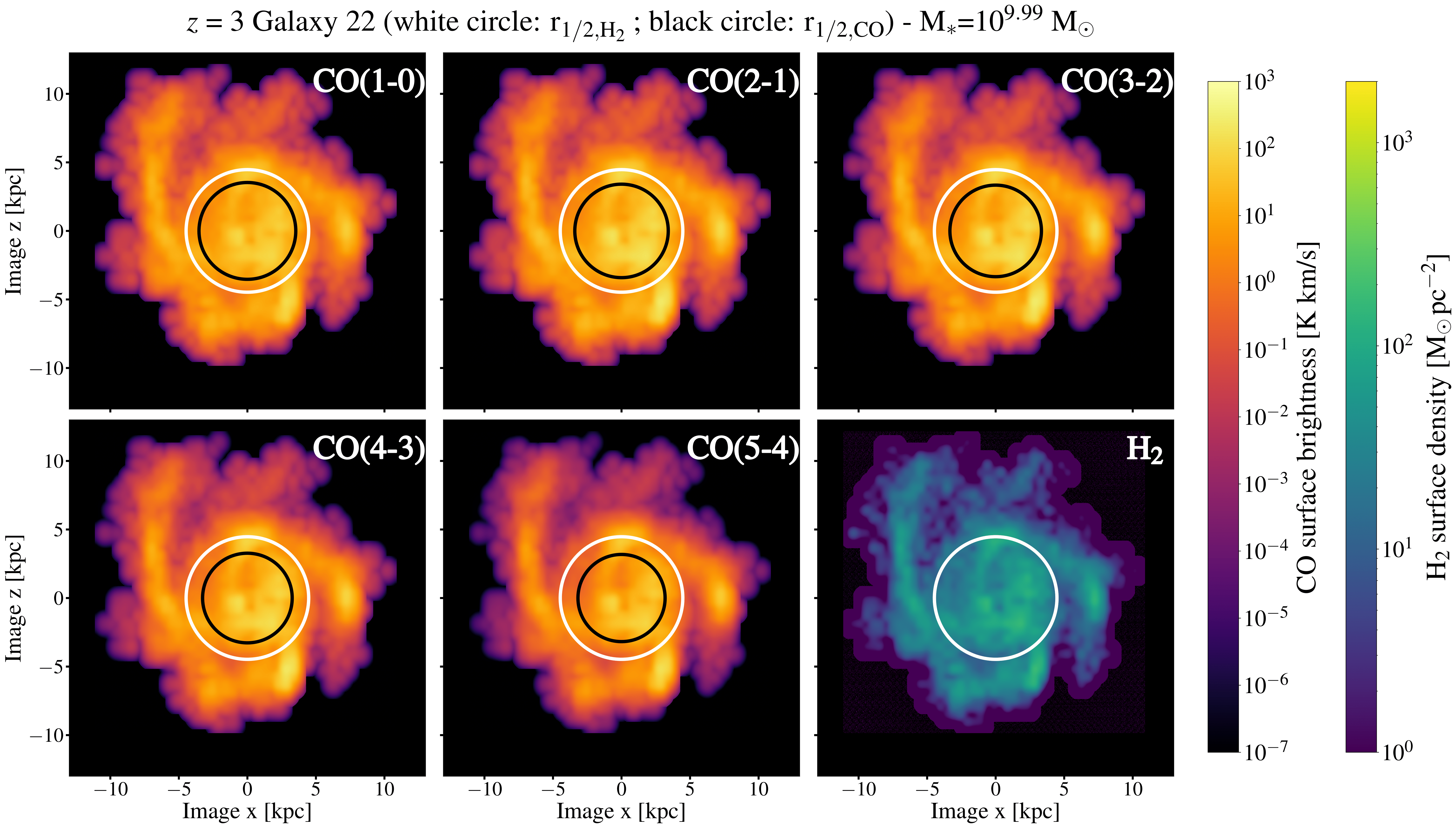}
        \caption{Example of the resolved CO emission and H$_2$ surface density from two massive galaxies, at $z=1$ (top) and $z=2.5$ (bottom). The half-light radii of the CO transitions labelled at the top of each panel are represented by the black circles, whereas the $\mathrm{H_2}$ half-mass radii are represented by the white circles. \label{fig:sample_emission}}
    \end{figure*}

    \subsection{Derived quantities}

        \SLICK\ calculates cloud properties for the entire three-dimensional gas structure of \SIMBA\ galaxies. To ensure the results are easily interpretable, we created 2D face-on projections, integrated along a viewing axis (z). We used the galaxy center of mass from \CAESAR\ and choose the viewing axis (normal vector) to be the resultant angular momentum vector of all gas bound to the galaxy (as identified by \CAESAR). Although this projection is biased toward obtaining extended disks and lower CO surface brightness, it aids in understanding the radial profiles discussed in Section \ref{subsec:radial}.
        
        For each 2D map, we used a constant ``pixel'' area of $500\times500$\,$\mathrm{pc^2}$. The line luminosities and molecular gas masses for clouds corresponding to a pixel position (assuming they are point-like) are summed up and divided by the pixel surface area to obtain each pixel's CO surface brightness and $\mathrm{H_2}$ surface density. We took medians for quantities such as the gas temperature, number density, and UV radiation field.

        In Fig.~\ref{fig:sample_emission}, we show a few examples of the CO surface brightness and H$_2$ surface density maps predicted using \SLICK\ for the $z$=1--3, main-sequence galaxies selected from \SIMBA. The white circles show the radius enclosing half the total mass of H$_2$ ($\mathrm{r_{1/2, H_2}}$), whereas the black circles show the CO half-light radii ($\mathrm{r_{1/2,CO}}$). For the massive galaxy (top panel), these velocity-integrated intensity maps show that the higher-$J$ CO emission is more pronounced in the regions with the highest H$_2$ surface density, namely in the spiral structures and nuclear ``ring'', dropping off significantly beyond these regions. In contrast (and as expected), the CO(1--0) emission better traces the diffuse emission at larger scales.

        We measured the sizes of galaxies in CO emission and $\mathrm{H_2}$ mass by calculating half-light and half-mass radii. These quantities are calculated by creating a face-on projection of the galaxy (as in the previous subsection) and estimating the radius from the \CAESAR\ center-of-mass to the gas particle, whose enclosed area contains half of the total CO luminosity or total $\mathrm{H_2}$ mass (see black and white circles in Fig.~\ref{fig:sample_emission}), respectively.

        Furthermore, we created radial profiles of CO emission for all galaxies by binning the pixels in radial bins of 0.5 kpc from a galaxy's \CAESAR\ center-of-mass up to 25 kpc. We did not consider pixels beyond 25 kpc from the center of the galaxy as nearly all of them have very low gas densities. We also excluded clouds for which \Despotic\ does not achieve chemical convergence (typically due to extremely low number densities).

\section{Results}
    \label{sec:results}

    \subsection{CO surface brightness vs H$_2$ surface density}

        \begin{figure*}
            \centering
            \includegraphics[width=0.9\textwidth]{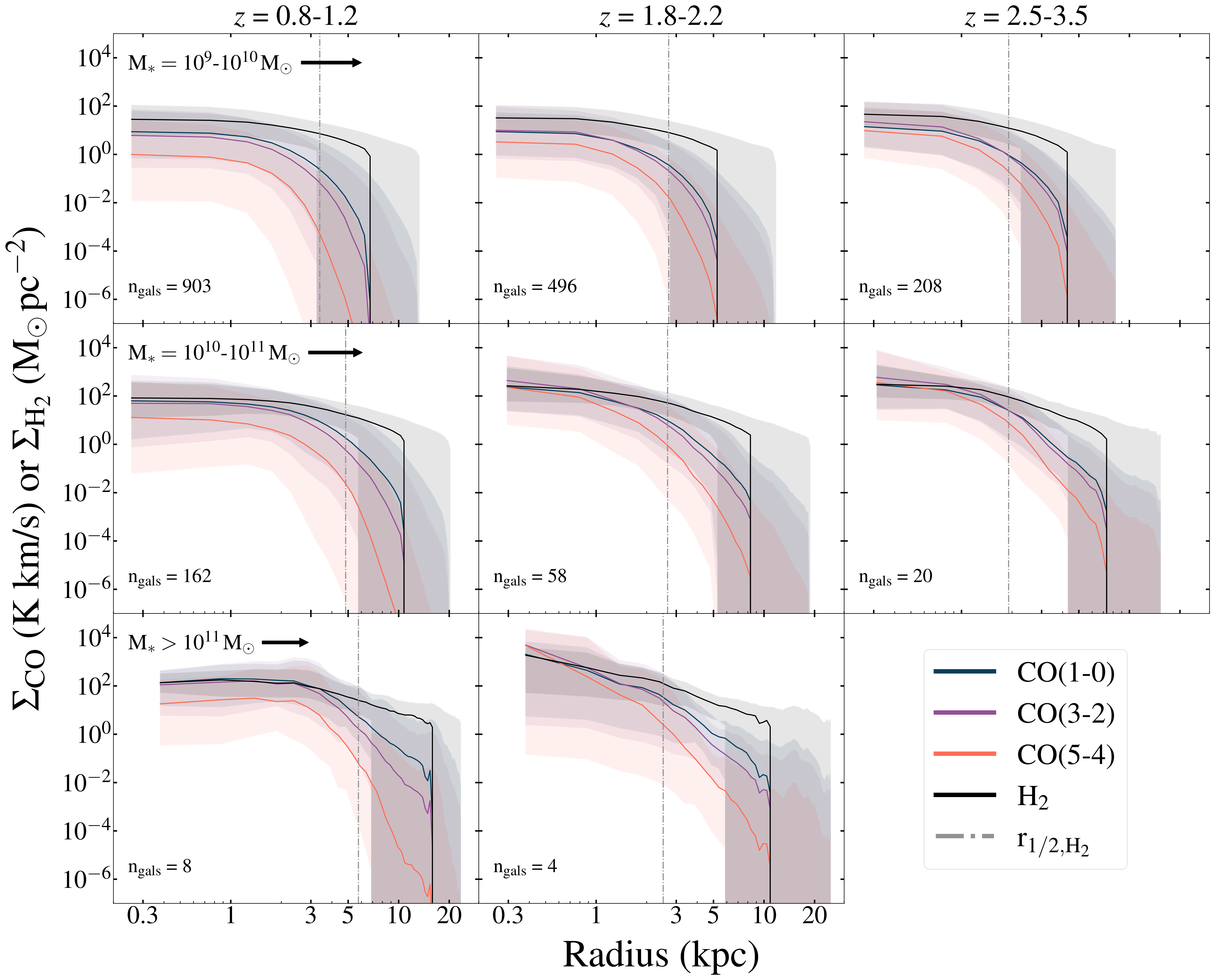}
            \caption{Average radial profiles of CO surface brightness and H$_2$ surface density. The x axis is plotted in log scale. The solid lines represent the median values per radial bin, averaged across the galaxy sample shown in each panel. The shaded regions encompass the 16th and 84th percentiles of the values per radial bin. Each column shows a different redshift slice, as labelled at the top, whereas each row is for a different stellar mass bin, as indicated in the left of each row. Given the volume of \SIMBA25, the highest mass bin contains the fewest galaxies. Only three CO transitions are shown, for clarity (legend at right).  \label{fig:COradialprofile}}
        \end{figure*}
    
        \begin{figure*}
            \centering
            \includegraphics[width=0.9\textwidth]{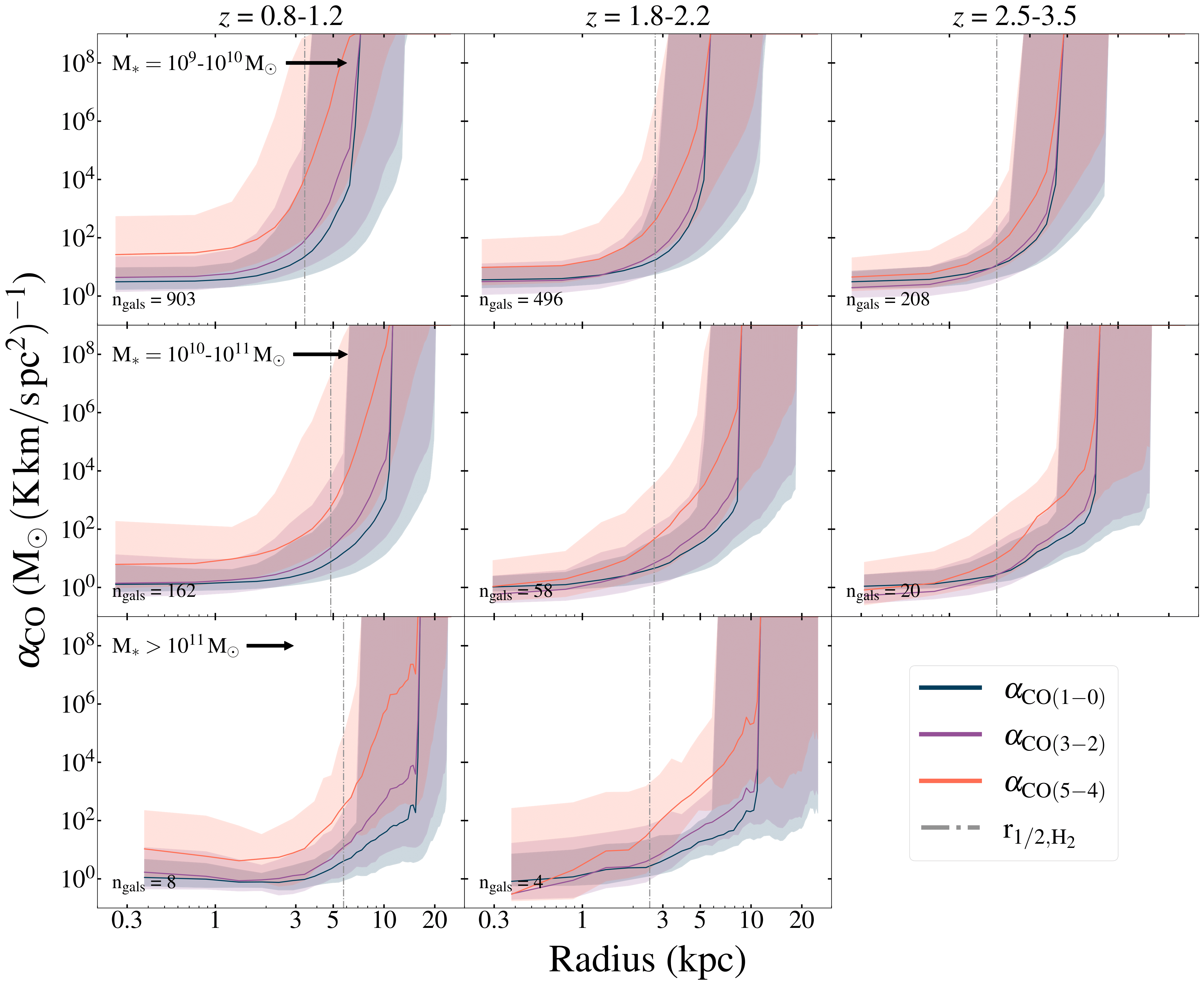}
            \caption{Average radial profile of the CO-to-H$_2$ conversion factor ($\alphaCO$), for different transitions (shown in the legend on the right). The x axis is plotted in log scale. The solid lines represent the median $\alphaCO$, averaged across the entire galaxy sample shown in each panel. The shaded regions encompass the 16th and 84th percentiles of the values at each radius. The columns and rows indicate the same redshift, and stellar mass bins as in Fig.~\ref{fig:COradialprofile}.  \label{fig:alphaCOradialprofile}}
        \end{figure*}

        We first test how well CO traces H$_2$ as a function of galaxy radius.  To this end, we perform a statistical test, determining the median (and 16th--84th percentiles) of the CO surface brightness per radial segment (defined in absolute radius) for galaxies in each redshift and stellar mass bin. Likewise, we determine the median (as well as the 16th and 84th percentiles in) H$_2$ surface density within each radial segment, for each redshift-stellar mass bin. In both cases, we consider pixels in the respective radial segments across all galaxies in the specified redshift and stellar mass bin. We compare the CO surface brightness profiles directly to the H$_2$ surface density profile in Fig.~\ref{fig:COradialprofile}. In Fig.~\ref{fig:alphaCOradialprofile}, we instead show the median (and 16th--84th percentile) radial profiles in the CO($J\to J-1$)-to-H$_2$ conversion factor, $\alphaCO$, which we again determine by taking the median $M_\mathrm{H_2} / L_\mathrm{CO(J\to J-1)}^\prime$ per radial segment, averaging over all galaxies within each redshift and stellar mass bin, for each radius. We show the results in terms of an absolute radius, noting that there is some variation in galaxy sizes per bin. We also tested normalizing the absolute radii by various ``galaxy radii'' (the largest distance at which a particle is identified, the half-H$_2$ mass radii, or half-light radii), but in each case the additional normalization added small additional trends that complicated the interpretation. 

        We find that for all bins of redshift and stellar mass, the CO surface brightness profiles decline more steeply than the H$_2$ surface density. This is also reflected in a sharp rise in $\alpha_\mathrm{CO(J \to J-1)}$ with radius (Fig.~\ref{fig:alphaCOradialprofile}), especially up to $\sim10$\,kpc. Thus, on average, the CO emission becomes fainter with respect to the H$_2$ density, with increasing radius---meaning that $\alphaCO$ increases significantly from $\sim$1 in the centers to $>$100 $\mathrm{M_{\odot}\,(K\,km/s\,pc^2)^{-1}}$ in the outermost regions. At lower redshifts, $\alpha_\mathrm{CO(J \to J-1)}$ increases more steeply, when matching in stellar mass. We also see weak trends with stellar mass; for galaxy populations with higher stellar mass, the rise in $\alphaCO$ is less steep, due to a combination of both the larger size and difference in physical conditions. The H$_2$ mass density profile drops off at $\sim$4--6 kpc for the low-mass galaxies but falls down to zero only at $\sim$15 kpc for the most massive galaxies as they tend to be more extended. We isolate the physical conditions driving these trends in Sec.\ref{subsec:alphaCO_disc}.
        
        In general, the CO surface brightness declines more steeply (and $\alpha_\mathrm{CO(J \to J-1)}$ rises more sharply) for higher $J$ transitions; CO(5--4) appears to be a weaker tracer of H$_2$ in the outskirts of these galaxies ($\gtrsim$5 kpc) than CO(1--0). The difference between the CO transitions is most pronounced for the lowest redshift ($z$=0.8-1.2) bin, where the CO(5--4) surface brightness in the outskirts is $10^4\,\times$ lower than for the $J<3$ CO transitions. Likewise, the surface brightness of the CO(1--0) and CO(3--2) transitions is more closely matched in the centers of the $z$=2.5--3.5 galaxy populations, than in the $z$=0.8--1.2 galaxies.

    \subsection{Radial variations in CO excitation}
        \label{subsec:radial}

        \begin{figure*}
            \centering
            \includegraphics[width=0.9\textwidth]{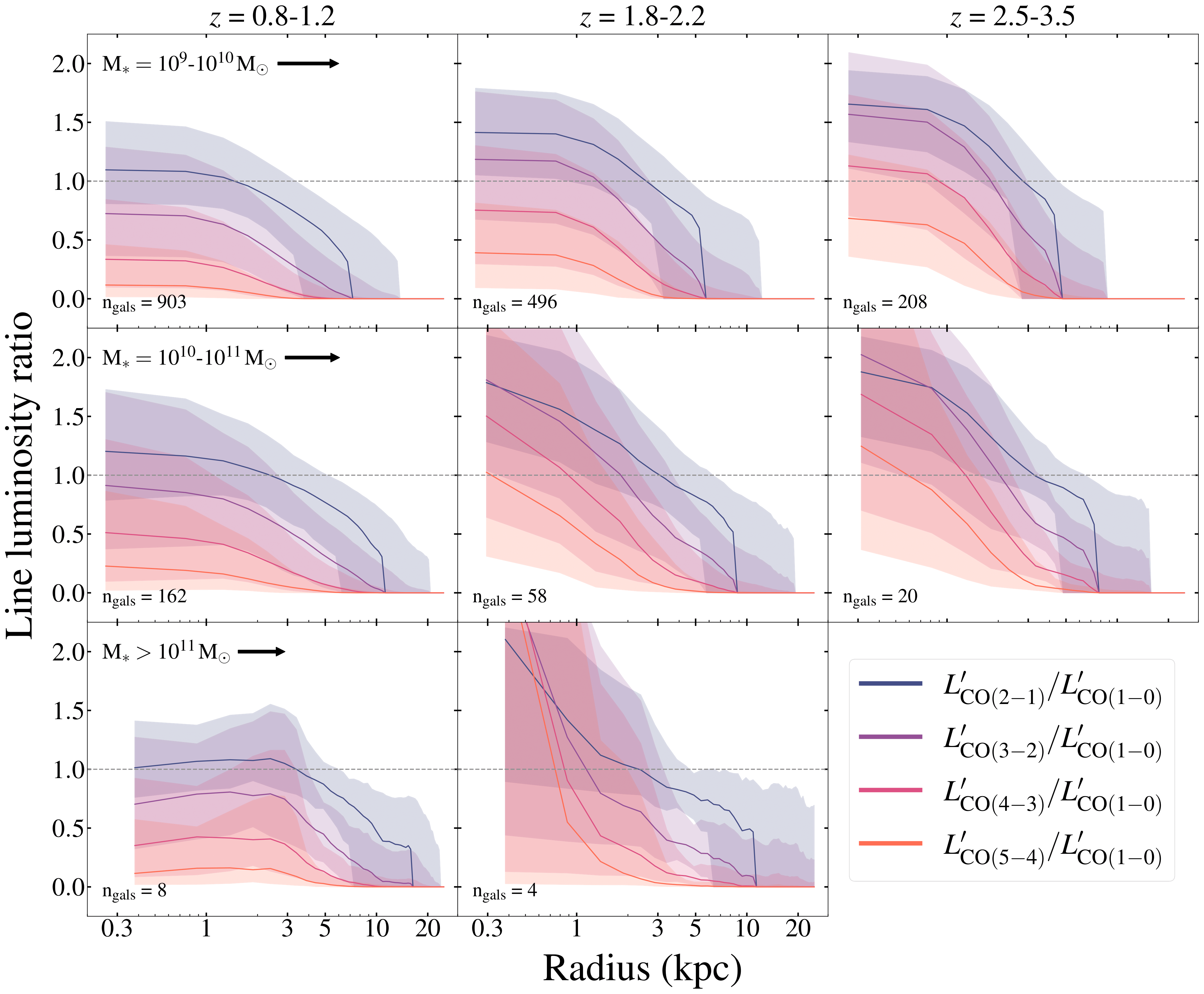}
            \caption{Average radial profile of the CO line luminosity ratios, in the same bins of redshift (columns) and stellar mass (rows) as the figures above. The solid lines represent the median line ratios, averaged across the entire galaxy sample shown in each panel, while the spread encompasses the 16th to 84th percentile values.\label{fig:lineratioradialprofile}}
        \end{figure*}

        To trace the CO excitation across these disks, we measure the radial profiles of the CO line luminosity ratios, relative to CO(1--0). We show the average radial profiles of these line luminosity ratios in Fig.~\ref{fig:lineratioradialprofile}. For all galaxy bins, we find that the line ratios peak in the center, drop at intermediate radii, and then flatten. We see a pronounced difference between different redshift bins; the values at both the centers and outskirts increase significantly with redshift and even exceed unity, such that the high-$J$ CO emission is suprathermal. For example, for the middle stellar mass range ($M_* = 10^{10}-10^{11}$), the CO(3--2)-to-CO(1--0) line ratio ($r_{31}$) peaks at 0.9, 1.7, and 2 for $z$=0.8--1.2, $z$=1.8--2.2, and $z$=2.5--3.5, respectively. We also find an increase in the line ratios (within the central regions) with stellar mass for redshifts$>$1.8, leading to a steeper decline in the line ratios for massive systems. Interestingly, the stellar mass trend is unclear in the lowest redshift bin, as AGN feedback begins quenching the centers of most star-forming galaxies by this time \citep{Dave_2020}. The trend at higher redshift may be related to the difference in sizes and hence central gas densities. We explore this further in Sec.~\ref{subsec:COexcitation_disc}.

    \subsection{CO sizes}

         \begin{figure}
            \centering
            \includegraphics[width=0.5\textwidth]{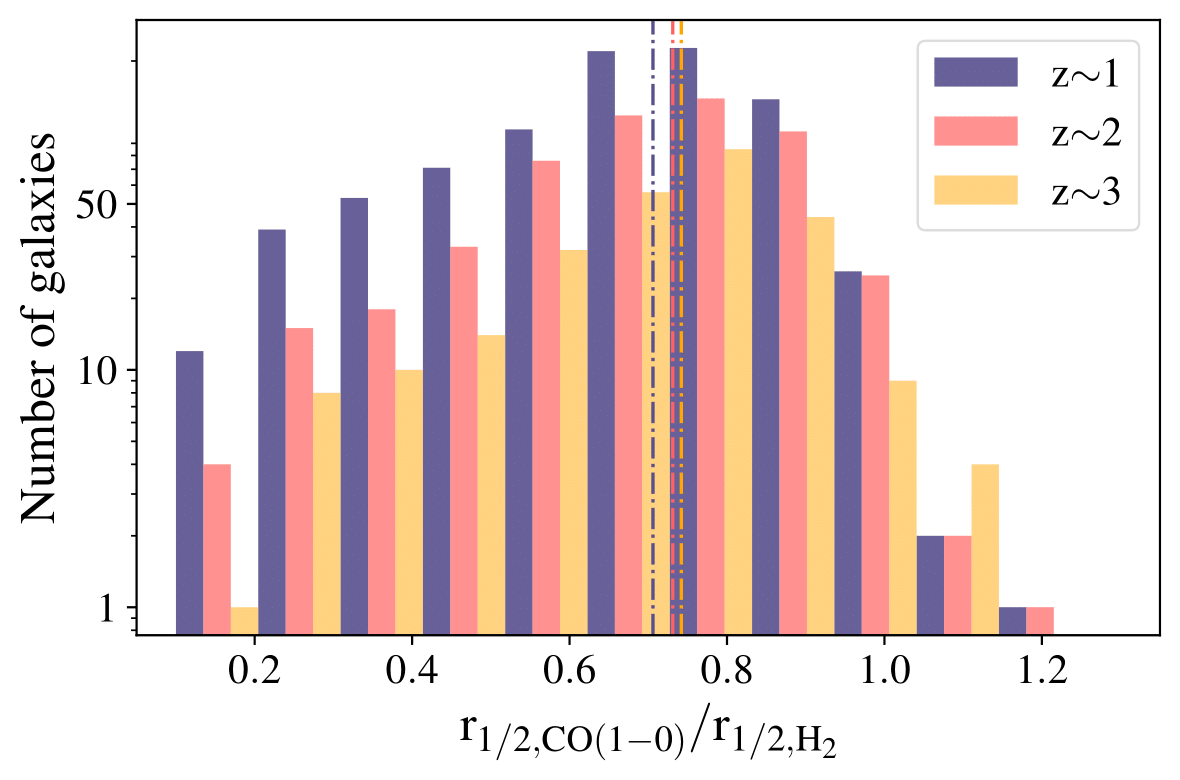}
            \caption{Histogram of the size ratio of CO(1--0) half-light radii ($\mathrm{r_{1/2,CO(1-0)}}$) to $\mathrm{H_2}$ half-mass radii ($\mathrm{r_{1/2,H_2}}$) for galaxies in 9 snapshots color-coded by three redshift bins (as shown in legend in the top right). The y axis is in log scale (with linearly described ticks). The dashed lines show the median value of the size ratio quoted in Table \ref{tab:stats}.} \label{fig:COhalfradii_over_H2halfmassradii}
        \end{figure}

        \begin{figure}
            \centering
            \includegraphics[width=0.4\textwidth]{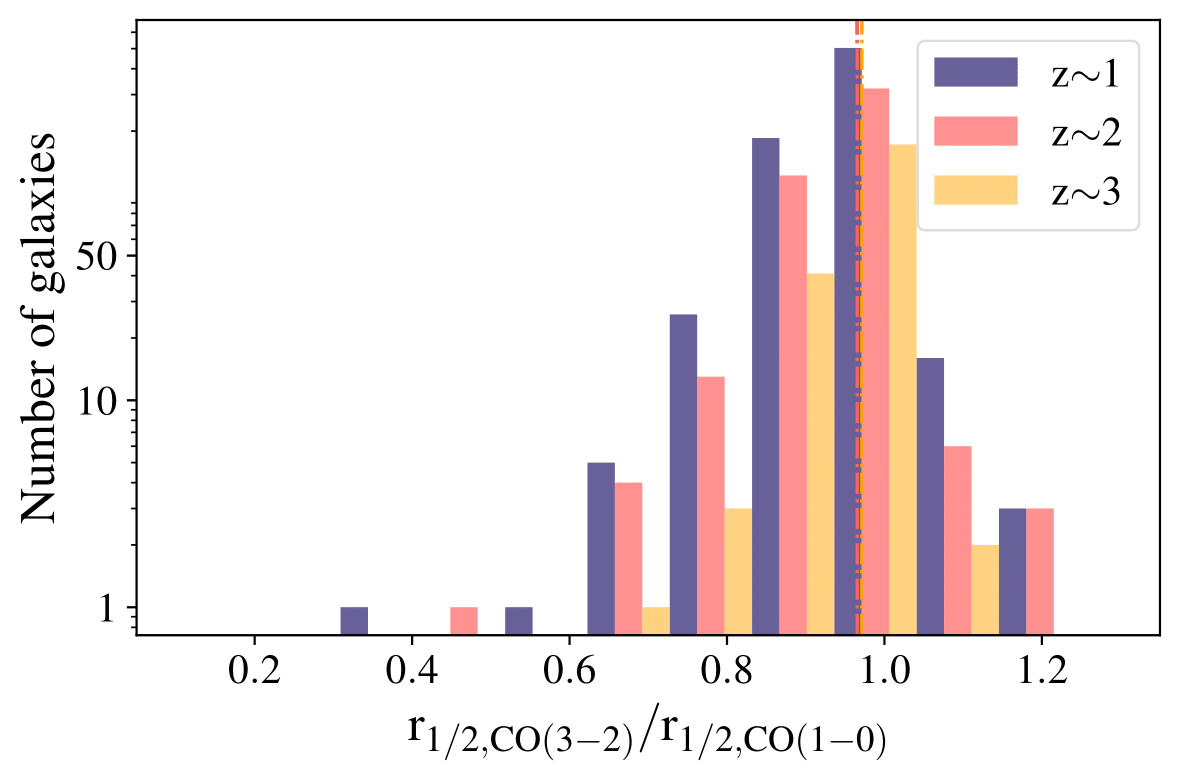}
            \includegraphics[width=0.4\textwidth]{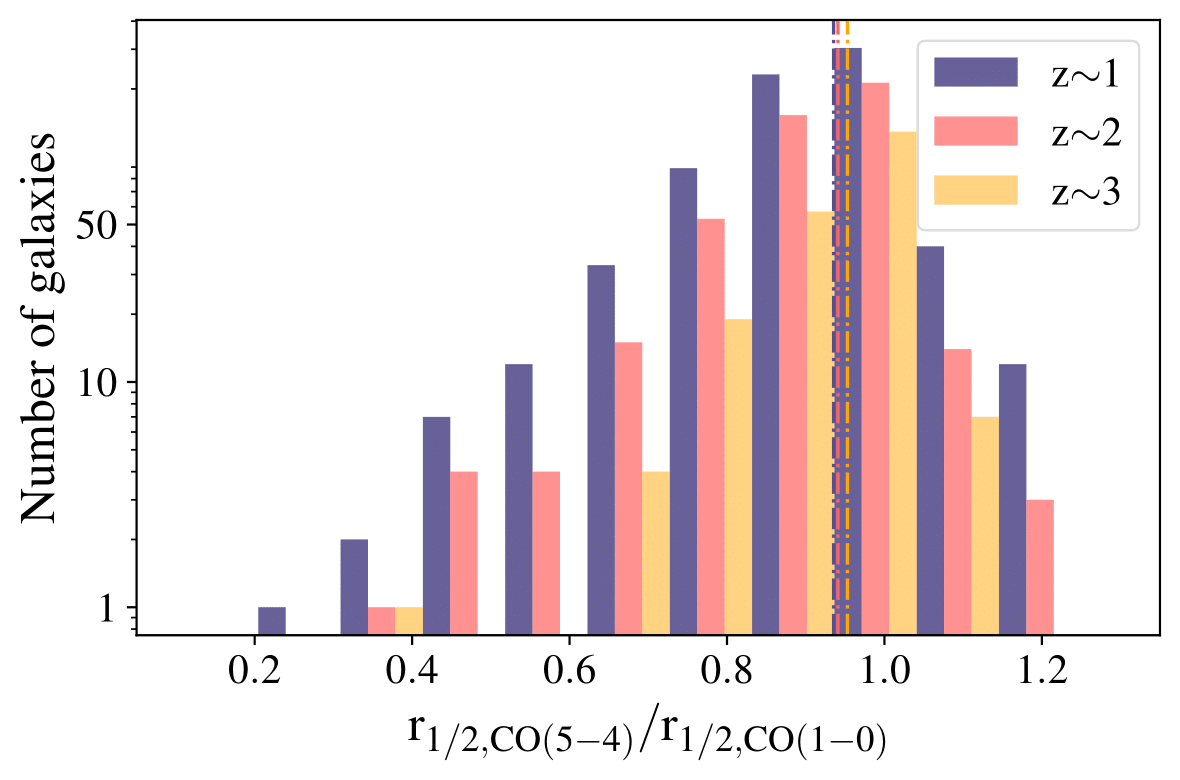}
            \includegraphics[width=0.4\textwidth]{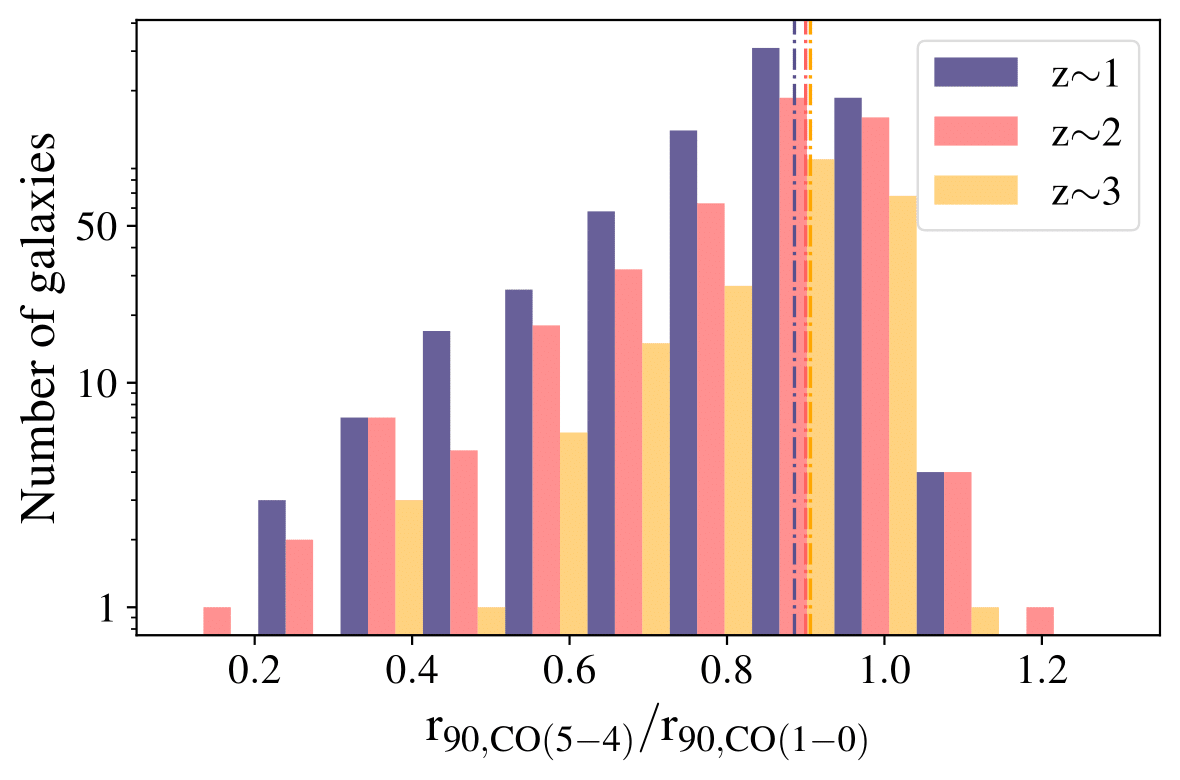}
            \caption{Histograms of the size ratios of (a) CO(3--2) half-light radii ($\mathrm{r_{1/2,CO(3-2)}}$) to CO(1--0) half-light radii ($\mathrm{r_{1/2,CO(1-0)}}$), (b) CO(5--4) half-light radii ($\mathrm{r_{1/2,CO(5-4)}}$) to CO(1--0) half-light radii ($\mathrm{r_{1/2,CO(1-0)}}$), and (c) CO(5--4) 90\%-light radii ($\mathrm{r_{90,CO(5-4)}}$) to CO(1--0) 90\%-light radii ($\mathrm{r_{90,CO(1-0)}}$) for galaxies in 9 snapshots color-coded by three redshift bins (as shown in legend in the top right). The y axis is in log scale (with linearly described ticks). The dashed lines show the median values of the size ratios quoted in Table \ref{tab:stats}.}
            \label{fig:COhighJ_over_CO10_halfradii}
        \end{figure}

        \begin{figure*}
            \centering
            \includegraphics[width=0.95\textwidth]{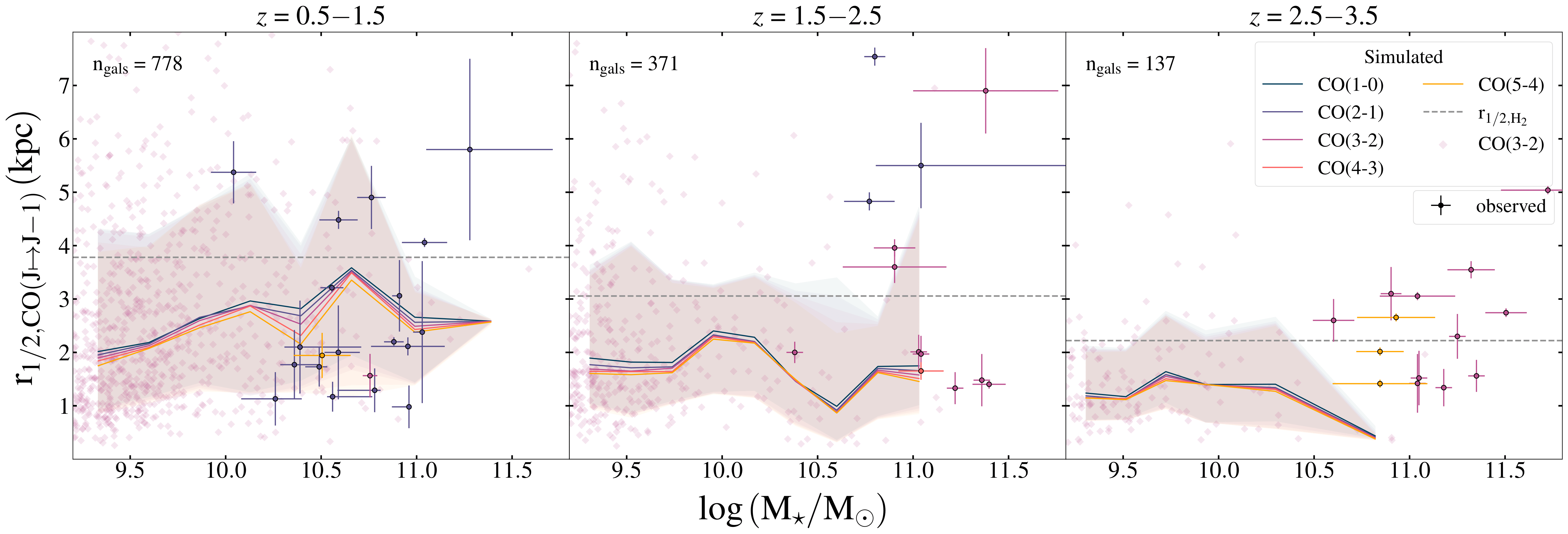}
            \caption{CO half-light radii as a function of galaxy stellar mass, in three redshift bins (one for each panel, as labelled at top), for both the simulated galaxies (line and shaded region) and observed galaxies. Median values are shown as solid lines, whereas the spread; that is, 16th--84th percentiles, are shown as shaded regions. Given the similarity between the half-light radii of the different CO transitions, we only show $\mathrm{r_{1/2,CO(3-2)}}$ of all galaxies as diamonds for clarity (see legend at right). Our predictions are compared to CO sizes measured from observations by \citealt{Kaasinen_2020,Ikeda_2022,Rizzo_2023,Tadaki_2023}.
            \label{fig:COhalfradii}}
        \end{figure*}

        \renewcommand{\arraystretch}{1.4}
        \begin{table}
        \begin{center}
        \caption{Size ratios and $\mathrm{H_2}$ fraction} 
        \begin{tabular}{@{}lcccc@{}}
            \toprule
               Quantity & $z$ = 0.8--1.2 & $z$ = 1.8--2.2 & $z$ = 2.5--3.5 \\
            \midrule
            \tablefootmark{a} $\mathrm{r_{1/2,CO(3-2)}}/\mathrm{r_{1/2,CO(1-0)}}$ & $0.96_{-0.06}^{+0.04}$ & $0.96_{-0.06}^{+0.03}$ & $0.97_{-0.06}^{+0.03}$ \\ \\
            \tablefootmark{b} $\mathrm{r_{1/2,CO(5-4)}}/\mathrm{r_{1/2,CO(1-0)}}$ & $0.93_{-0.12}^{+0.08}$ & $0.93_{-0.11}^{+0.06}$ & $0.95_{-0.11}^{+0.05}$ \\ \\
            \tablefootmark{c} $\mathrm{r_{1/2,CO(1-0)}}/\mathrm{r_{1/2,H_2}}$ & $0.69_{-0.21}^{+0.15}$ & $0.72_{-0.18}^{+0.14}$ & $0.74_{-0.2}^{+0.1}$ \\ \\
            \tablefootmark{d} H$_2$ fraction @ & & \\
            $r <= 2\mathrm{\,r_{1/2,CO(1-0)}}$ & $0.7_{-0.2}^{+0.2}$ & $0.72_{-0.19}^{+0.17}$ & $0.7_{-0.19}^{+0.17}$ \\
                \bottomrule
        \end{tabular}
        \tablefoot{\tablefoottext{a}{CO(3--2) half-light to CO(1--0) half-light radii ratio}, \tablefoottext{b}{CO(5--4) half-light to CO(1--0) half-light radii ratio}, \tablefoottext{c}{CO(1--0) half-light to $\mathrm{H_2}$ half-mass radii ratio}, and \tablefoottext{d}{ H$_2$ mass fraction within twice the CO(1--0) half-light radius}, for galaxies in three redshift bins. The median values are shown here as well as the 16th \& 84th percentiles as sub/super-scripts respectively. \label{tab:stats}}
        \end{center} 
        \end{table} 
        \renewcommand{\arraystretch}{1}

        We test whether the extent of the CO emission is compatible with that of the H$_2$ mass.
        Fig.~\ref{fig:COhalfradii_over_H2halfmassradii} shows the ratio of the CO(1--0) half-light radius vs. H$_2$ half-mass radius ($\mathrm{r_{1/2,CO(1-0)}/r_{1/2,H_2}}$) for the three redshift bins. 
        In the vast majority of cases, the CO half-light radii are more compact than the H$_2$ half-mass radii (i.e., $\mathrm{r_{1/2,CO(1-0)}/r_{1/2,H_2}}$), as indicated by the median half-light to half-mass ratios given in Table~\ref{tab:stats}. Only a handful have CO half-light radii in excess of the H$_2$ half-mass radius. We visually inspected these and find that this scenario arises from infalling clumps of gas that are much brighter in CO(1--0) than in the center of the galaxy they are assigned to. We find that the median $\mathrm{H_2}$ fraction contained within twice the CO(1--0) half-light radius is 0.7, 0.72, 0.7 for $z$=0.8--1.2, $z$=1.8--2.2, and $z$=2.5--3.5, respectively (see Table~\ref{tab:stats}). Thus, on average, more than 25\% of the $\mathrm{H_2}$ mass is located outside $2\mathrm{\,r_{1/2,CO(1-0)}}$. We find a slight increase in the median size ratio with redshift.

         We further compare the sizes of the different CO transitions in Fig.~\ref{fig:COhighJ_over_CO10_halfradii} using two metrics: $\mathrm{r_{1/2}}$, which is commonly used in $z>1$ observational studies, and $\mathrm{r_{90}}$; that is, the radius within which 90\% of the light is contained. For most galaxies, the higher-$J$ transitions are more compact than the CO(1--0) emission, tracing the sharp decline in line luminosity ratios with galactocentric radius. The median ratios between the half-light radii of CO(5--4) vs. CO(1--0) are 0.93, 0.93, 0.95 for $z$=0.8--1.2, $z$=1.8--2.2, and $z$=2.5--3.5, respectively (see Table~\ref{tab:stats}). We find a similar median $\mathrm{r_{1/2,CO(3-2)}/r_{1/2,CO(1-0)}}$ ratio of 0.96, 0.96, 0.97 for the three redshift bins. The difference in sizes becomes more pronounced if we instead use $\mathrm{r_{90}}$, as less of the diffuse emission is traced by the higher-$J$ transitions (see Fig.~\ref{fig:COhighJ_over_CO10_halfradii}).

        Next, we place the half-light radii of the different CO transitions measured here in context with observed galaxies (Fig.~\ref{fig:COhalfradii}). Despite the clear difference in the sizes and surface brightness profiles for different transitions, we find a strong decrease in the half-light radii with increasing $J$ only for the lowest redshift bin. We note that the galaxy-to-galaxy scatter is large, due to the diversity of relative compactness of the high-$J$ to CO(1-0) sizes (as seen in Fig.~\ref{fig:COhighJ_over_CO10_halfradii}). For the most part, the half-light radii are very compact, at 1--5\,kpc, consistent with most $z$=1--3 star-forming galaxies observed to date. These (median) values are significantly smaller than the radius at which the H$_2$ surface density or $\alphaCO$ profiles flatten, again implying that most observations to date have only been sensitive to these central regions.
        
        For the lowest redshift bin, we find that that the CO half-light radii measured from observations are systematically lower than for our simulated sample; whether this is due to the observational bias toward compact, submillimetre-bright galaxies, sensitivity plus measurement biases in observations, or an overprediction of the simulated galaxy sizes due to the sub-resolution modeling is challenging to test without a sufficient sample of massive simulated galaxies. We also clearly recover an expected increase in the CO half-light radii with stellar mass. However, we do not recover the same trend for the higher redshift bins. We caution that despite some systematic differences between observations and our median trends, individual galaxies can show much smaller or larger half-light radii (see diamonds).
        
        We investigate what might cause the compact sizes further in Sec.~\ref{sec:discussion}, noting already that the gas reservoirs are clearly centrally concentrated and that most of the densest gas lies within the central 5\,kpc. Thus, the half-light radii of all CO transitions are relatively compact and change little with the rotational transition, despite marked differences further out. We also note that while the averaged results show little difference, there is a huge scatter about the half-light radii measurements, with large differences between the measurements for different transitions, for individual galaxies. Thus, we caution against comparing these sample averaged predictions to any one observed galaxy.

\section{Discussion}
    \label{sec:discussion}
    
    \subsection{The physical conditions driving the radial decline in $\alphaCO$}\label{subsec:alphaCO_disc}

        \begin{figure*}
            \centering
            \includegraphics[width=0.9\textwidth]{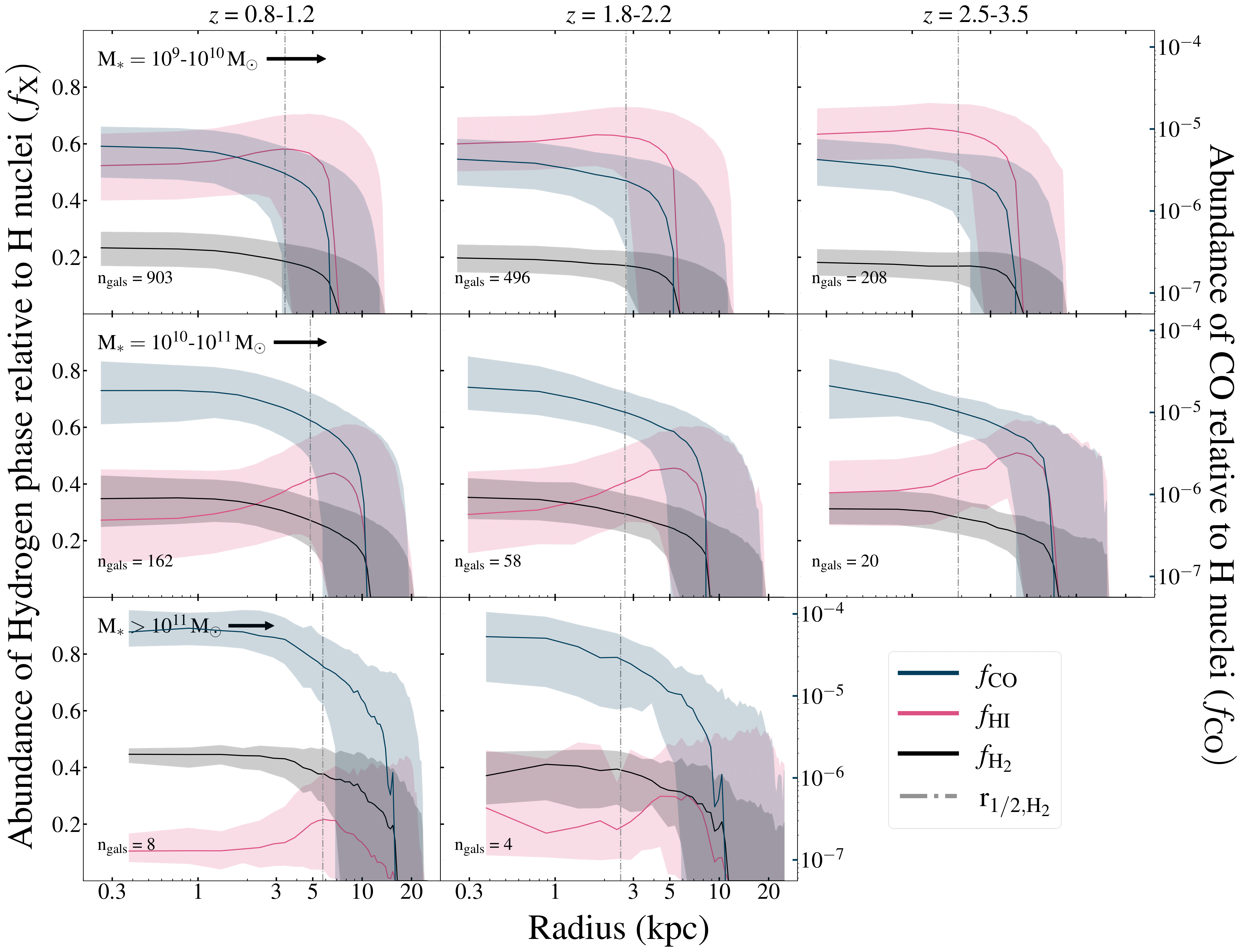}
            \caption{Average radial profile of abundances of HI, $\mathrm{H_2}$ (left axis), and CO (right axis) per H nucleus. The profiles are shown in the same bins of redshift (columns) and stellar mass (rows) as Figs.~\ref{fig:COradialprofile}--~\ref{fig:lineratioradialprofile}. The solid lines represent the median relative abundances, averaged across the entire galaxy sample corresponding to each panel, while the shaded regions indicate the 16th to 84th percentiles of the values at each radial bin. The three species are color-coded as shown in the legend at the bottom right. Note that the maximum possible value for HI would be 1 (i.e., fully atomic), but for $\mathrm{H_2}$ would be 0.5 (i.e., fully molecular). \label{fig:abundance_profiles}}
        \end{figure*}

        One of our main results is the steep decrease in CO surface brightness beyond the central few kiloparsecs (and beyond the $\mathrm{H_2}$ half-mass radii) of galaxies across stellar mass and redshift ranges. This is naturally accompanied by the steep increase in $\alphaCO$ with galactocentric radius, as the $\mathrm{H_2}$ mass surface density drops less sharply than the CO luminosity. Increases in $\alphaCO$ toward galaxy outskirts have also been observed for nearby star-forming galaxies \citep[e.g.][]{Sandstrom_2013,Yasuda_2023,Chiang_2024}, where the decline has been suggested to scale with lower velocity dispersions and ISRF strengths. Moreover, lower metallicities, temperatures, and/or gas densities are expected to drive increases in $\alphaCO$ \citep[e.g.][]{Bolatto_2013}.
        To understand what is driving these trends in \SIMBA\ and our forward modeling, we assess the impact on $\alphaCO$ of chemical abundance, hydrogen number density, metallicity, the ISRF, and the non-thermal velocity dispersion of the gas.
        
        We test how the CO, H$_2$, and \HI\ abundances per H nuclei vary with galactocentric radius in Fig.~\ref{fig:abundance_profiles}. We find that the fractional CO abundance declines sharply by up to one order of magnitude beyond the central few kiloparsecs. From this test, it is clear that the CO abundance plays a major role in the increase in $\alphaCO$ at large radii; that is, there is simply less CO per amount of H$_2$ that is able to emit. The CO abundance (see y axis on the right) declines far more sharply than the H$_2$ abundance in the galaxy outskirts, indicating the presence of more "CO-dark" molecular gas \citep{Madden_2020}. As a sanity check of our modeling, we show that the contribution of \HI\ increases with radius as the H$_2$ fraction decreases, as expected from observations of local galaxies \citep[e.g.][]{Schruba_2011}.
        
        We find that the CO and H$_2$ abundances increase with galaxy stellar mass (Fig.~\ref{fig:abundance_profiles}, different rows), as expected due to the increased metal enrichment in massive galaxies \citep{Dave_2020}. Similar analyses of local galaxies have also shown that $\alphaCO$ decreases with increasing stellar mass and stellar mass surface density, ($\mathrm{\Sigma_{*}}$), which in turn correlate with increasing metallicity \citep{Bolatto_2013,Chiang_2024}. Beyond 5--10 kpc, the \HI\ fraction also declines, and only ionized gas remains. Outside the disks, we do not recover any atomic or molecular hydrogen.
        
\begin{figure*}
            \centering
            \includegraphics[width=0.9\textwidth]{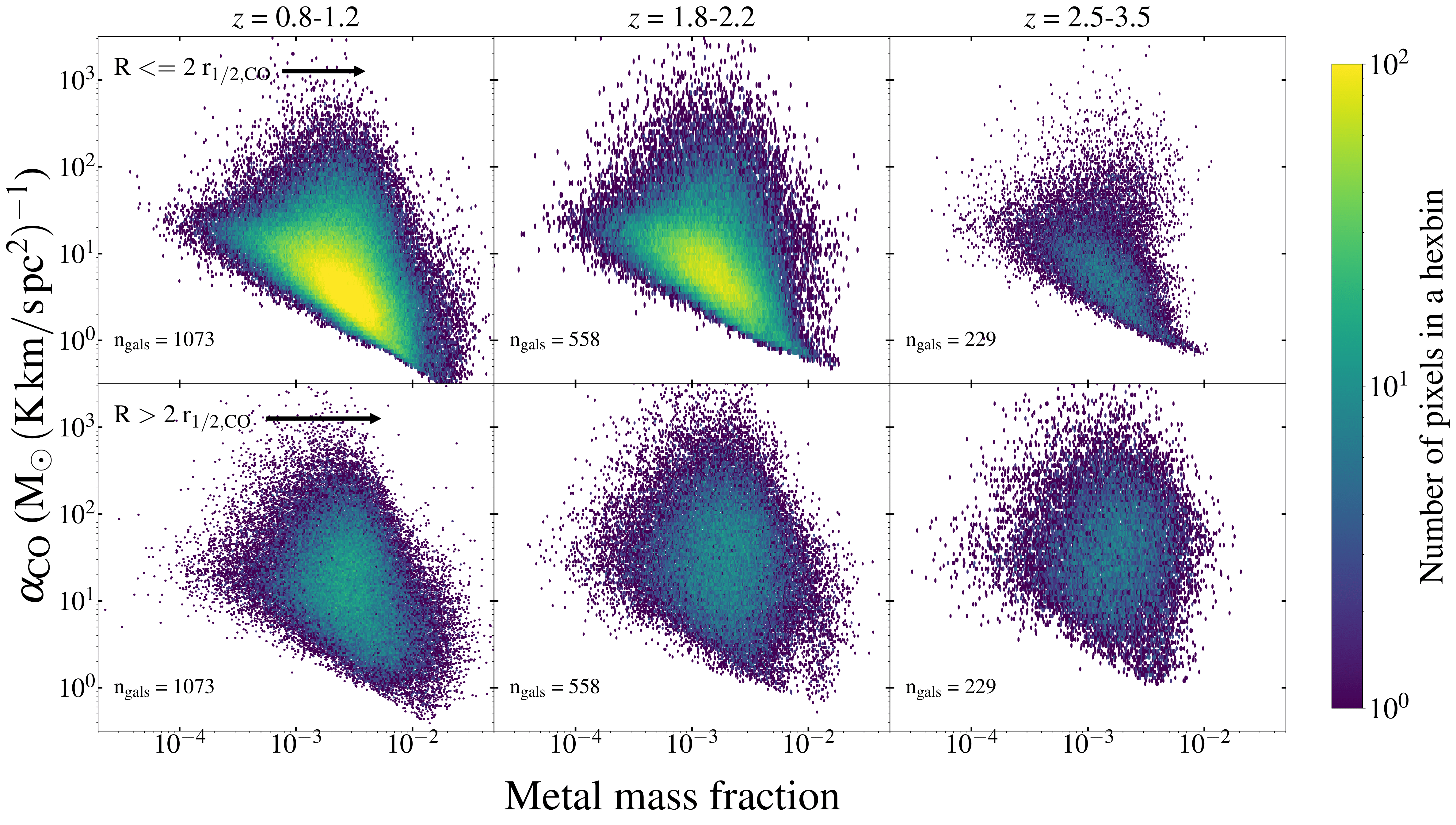}
            \caption{$\alphaCO$ vs. gas-phase metallicity per pixel, including the pixels of all synthetic CO(1--0) maps for the entire $z$=1--3 sample. The color-coding represents the number of pixels per hexbin as shown via a color bar on the right of the figure. The columns again show different redshift bins, whereas the rows indicate the radial position of the clouds within the galaxies with respect to twice the CO(1--0) half-light radius of the galaxy: inner (top), outer (bottom). \label{fig:alphaCOvsZ}}
        \end{figure*}

\begin{figure*}
            \centering
            \includegraphics[width=0.9\textwidth]{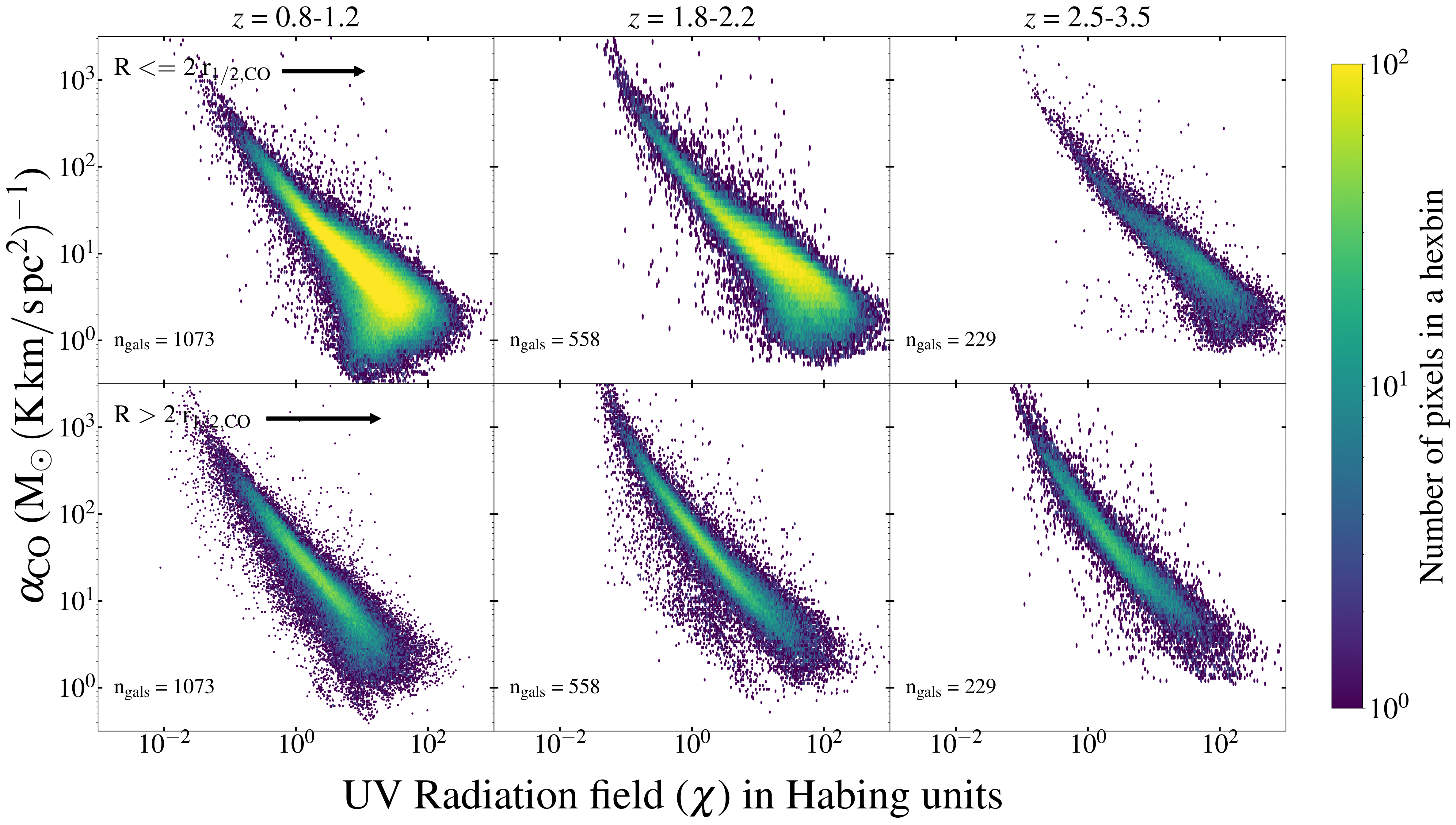}
            \caption{$\alphaCO$ vs. UV field strength, $\chi$, per pixel. The column and row separation, and color-coding are the same as for Fig.~\ref{fig:alphaCOvsZ}. \label{fig:alphaCOvschi}}
        \end{figure*}

\begin{figure*}
            \centering
            \includegraphics[width=0.9\textwidth]{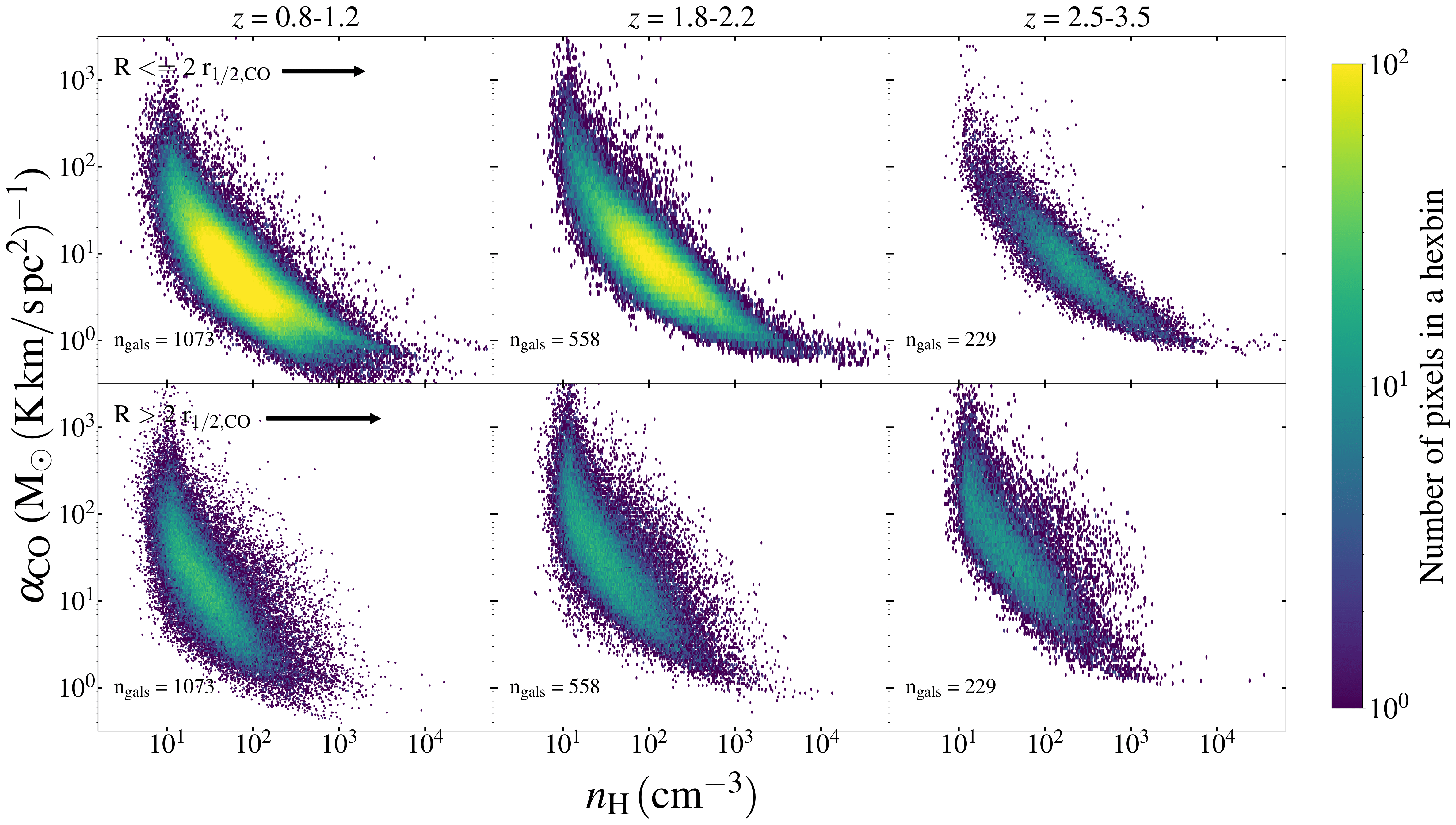}
            \caption{$\alphaCO$ vs. hydrogen number density per pixel. The column and row separation, and color-coding are the same as for Fig.~\ref{fig:alphaCOvsZ}.  \label{fig:alphaCOvsnH}}
        \end{figure*}

\begin{figure*}
            \centering
            \includegraphics[width=0.9\textwidth]{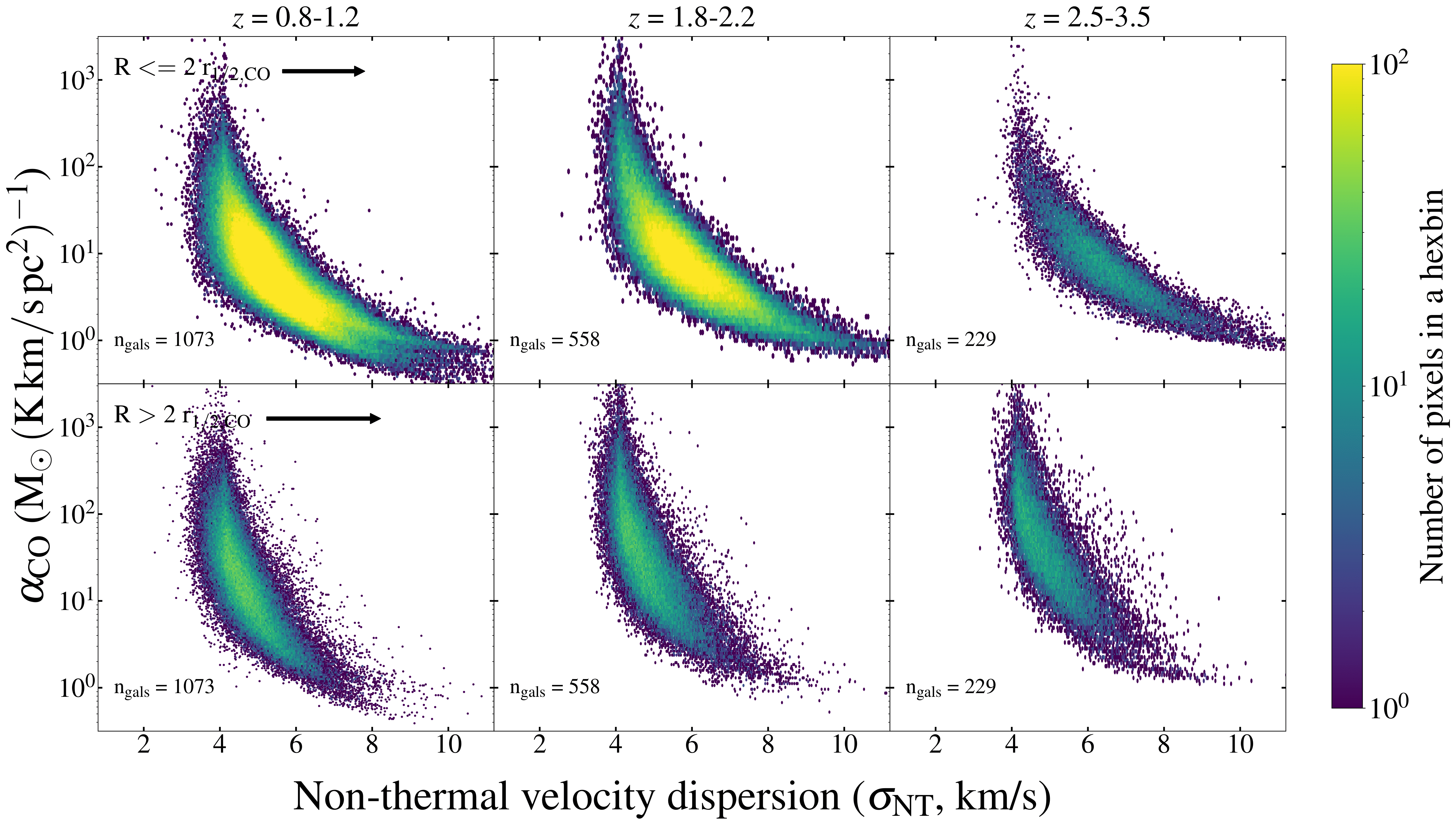}
            \caption{$\alphaCO$ vs. non-thermal velocity dispersion per pixel. The column and row separation, and color-coding are the same as for Fig.~\ref{fig:alphaCOvsZ}.  \label{fig:alphaCOvssigmaNT}}
        \end{figure*}

        We further investigate how the physical conditions within our sub-resolution cloud population regulate $\alphaCO$, by testing how $\alphaCO$ varies as a function of gas-phase metal fraction (Z), ISRF strength ($\chi$), hydrogen number density ($n_{\rm H}$), and non-thermal velocity dispersion ($\mathrm{\sigma_{NT}}$) respectively, in Figs.~\ref{fig:alphaCOvsZ}--\ref{fig:alphaCOvssigmaNT}. To this end, we derive the $\alphaCO$ of each pixel (see Sec.~\ref{sec:results}), by using the total $M_\mathrm{H_2}/L_\mathrm{CO(1--0)}^\prime$ of that pixel. Likewise, we determine the metal fraction, $\chi$, $n_{\rm H}$, and $\mathrm{\sigma_{NT}}$ of each pixel, where the color-coding represents the number of pixels in each ``hexbin''. Here, ``hexbin'' refers to the 2-dimensional hexagonal binning of points on the parameter space. To illustrate radial changes, we separate these pixels (in two rows) by their radial position in the galaxy, defined relative to twice the CO(1--0) half-light radius ($\mathrm{r_{1/2,CO}}$). As is clear from comparing the two rows of each figure, most of the CO emission stems from the inner part of galaxies (within twice the $\mathrm{r_{1/2,CO}}$), with very little CO emission in the outskirts.
        
        Although abundance plays an important role, Fig.~\ref{fig:alphaCOvsZ} reveals that changes in $\alphaCO$ are not driven solely by the metallicity (and its implicit regulation of the CO abundance). As expected \citep[see][and references therein]{Bolatto_2013}, the majority of pixels do follow an anticorrelation between $\alphaCO$ and the surface-averaged gas-phase metal fraction; the higher the metallicity, the more CO is self-shielded, preventing it from dissociating. However, for metal fractions between $\mathrm{(10^{-3}-10^{-2})}$, there is a significant fraction of pixels with higher $\alphaCO$, which do not follow this trend. These correspond to clouds with moderate metallicity that have very low number densities, and thereby produce lower CO(1-0) emission. This additional dependence on density has also been reported in local studies of low-metallicity galaxies. \citet{Hunt_2023} derive $X_{\rm CO}$ values 5--1000 times higher than the Milky-Way value for metal-poor starbursts. They find an order of magnitude scatter in the derived $X_{\rm CO}$ at 250 pc resolution compared to the expectation from the metallicity trend, pointing to a secondary dependence on the CO brightness temperature. Similarly, \citet{Ramambason_2024} find that the ``clumpiness'' of the gas in local star-forming dwarfs correlates with the dispersion in the $\alphaCO$-$Z$ relation.
        
        Fig.~\ref{fig:alphaCOvschi} indicates that there is a much stronger, linear anticorrelation between $\alphaCO$ and the ISRF strength (compared to $\alphaCO$ vs. metallicity). Moreover, there is a clear anticorrelation with density (Fig.~\ref{fig:alphaCOvsnH}); that is, $\alphaCO$ is insensitive to density in the lowest and highest-density regimes, but decreases with $n_{\rm H}$ in the range $n_{\rm H} \sim 20 - 300$ cm$^{-3}$. These effects are driven by the more efficient production of CO and $\mathrm{H_2}$ in denser gas, along with changes in the non-thermal velocity dispersion ($\sigma_{\rm NT}$) and gas temperature, both of which scale with the density and UV radiation field due to our sub-resolution modeling approach (Sec.~\ref{subsec:slick}).
        
        The clear anticorrelation between $\alphaCO$ and $\chi$ results from the ISRF increasing the kinetic temperature of the gas in regions where the density also remains high. Because the sources of the ISRF contribute proportionally to the CR field strength (Sec.~\ref{sec:method}), the gas ionization heating by cosmic rays is also enhanced in regions of large $\chi$. The higher $T_{\rm g}$ (caused by the increase in ISRF and CR strength), increases the brightness temperature and hence luminosity of CO(1--0), leading to a decrease in $\alphaCO$. These effects are additionally boosted by \Despotic's treatment of gas clumping; as $f_\mathrm{cl}$ increases (due to higher $\sigma_{\rm NT}$), CO is more efficiently excited. Since $\sigma_\mathrm{tot}$ and $n_{\rm H}$ are both inversely proportional to the cloud radius $R_{\rm C}$ (see Sec.~\ref{subsec:slick}), the velocity dispersions also increase with density. As $\sigma_{\rm tot}$ increases, the optical depth of the CO lines decreases, allowing more CO emission to escape the cloud.
        
        The combination of a high-density environment, strong $\chi$, and large $\sigma_{\rm NT}$ enables higher collisional excitations and CO self-shielding, while preventing $\mathrm{H_2}$ photodissociation. These regions are preferentially located in the inner parts of galaxies, within $\mathrm{2\times r_{1/2,CO}}$ (top row of Figs.~\ref{fig:alphaCOvschi}--\ref{fig:alphaCOvssigmaNT}), where there is active star formation. Therefore, pixels with the lowest $\alphaCO$ lie in the densest inner regions and spiral arms, which have high velocity dispersions and high ISRF strengths. The outer regions of our modeled galaxies exhibit severe CO depletion (and significant $\mathrm{H_2}$ photodissociation), accompanied by decreasing densities and $\chi$, leading to an increase in $\alphaCO$ with galactocentric radius. This increase begins at smaller radii for low-mass and/or high-redshift galaxies (see Fig.~\ref{fig:alphaCOradialprofile}), as the molecular interstellar medium (ISM) is restricted to the innermost regions. The compactness of the dense ISM across redshift is also indicated in the comparison of CO and $\mathrm{H_2}$ sizes in Fig.~\ref{fig:COhalfradii}.

        Our findings differ somewhat from the models of \citet{Olsen_2016}, who report that $\alphaCO$ stays fairly constant in the range of $\sim1.4-1.8$ in their three, massive ($0.5-2.1\times10^{11}\,\Msun$) simulated galaxies. They find that $\alphaCO$ only increases marginally beyond 5 kpc, due to the decrease in $G_0$ (equivalent to $\chi$ here). The increase in $\alphaCO$ in our models is steeper and reaches values of up to 70 (K km s$^{-1}$ pc$^2$)$^{-1}$  for galaxies at the same stellar mass and redshift, due to the stronger dependence on the CO abundance (and $\chi$). This sharper dependence with respect to the models of \citet{Olsen_2016} is likely driven by several differences in our models assumptions. \citet{Olsen_2016} assume a fixed, Milky-Way $\mathrm{[CO/H_2]}$ value of $2\times10^{-4}$ for dense gas ($n_{\rm H_2}>50\,\mathrm{cm^{-3}}$), whereas this value can be a factor of 2-100 lower in our galaxies (at the same densities). Moreover, our results provide an important distinction between metallicity and CO abundance. Namely, even though $\alphaCO$ is not strongly anti-correlated with the gas-phase metallicity, the CO abundance can drop by many orders of magnitude even if the metallicity does not, as CO is destroyed quite efficiently at low densities.

        The strong dependence of $\alphaCO$ on local ISM properties such as the ISRF and velocity dispersion in our models (Figs.~\ref{fig:alphaCOvschi} and \ref{fig:alphaCOvsnH}) is consistent with findings from observational studies of local galaxies \citep[e.g.][]{Sandstrom_2013,Yasuda_2023,Chiang_2024}. For instance, \citet{Chiang_2024} used FIR dust emission maps and an assumption on the dust-to-metal ratio to argue that $\alphaCO$ correlates most strongly with the SFR surface density and the local ISRF at 2 kpc scales. In our modeling, these properties are set indirectly by the \SIMBA\ galaxy formation physics through the cooling and star formation prescriptions (see Section \ref{subsec:slick}). Likewise, $\alphaCO$ is also lower in the interiors of galaxies, where the stellar surface densities are typically much higher.

    \subsection{The physical drivers of variations in CO excitation}\label{subsec:COexcitation_disc}

        In Sec.~\ref{subsec:radial}, we provided evidence for a marked decline in the CO excitation with radius and an overall increase with redshift, especially in the central regions. We test whether these trends are driven by changes in hydrogen number density ($n_{\rm H}$) and/or gas temperature ($T_{\rm g}$) in Figs.~\ref{fig:lineratiovsnH} and~\ref{fig:lineratiovsTg}, taking these values from the output of \Despotic. To simplify the visual interpretation, we only show trends for the CO(3--2)/CO(1--0) line ratio ($r_{31}$), noting that the other line ratios follow the same (albeit less or more enhanced) trends. Like in Fig.~\ref{fig:alphaCOvsZ}, we derive $r_{\rm_{31}}$, $n_{\rm H}$, and $T_{\rm g}$ for each pixel and bin these values. To illustrate radial changes, we bin these pixels (in two rows) by their radial position in the galaxy, defined relative to twice the CO(1--0) half-light radius ($\mathrm{r_{1/2,CO}}$). The columns are consistent with the previous figures, showing the same redshift bins.

        The increase in CO excitation toward the galaxy centers and with increasing redshift can be mostly attributed to an increase in hydrogen number density. Fig.~\ref{fig:lineratiovsnH} shows that $r_{31}$ increases substantially with $n_{\rm H}$. Moreover, it is clear that the central regions of galaxies ($\mathrm{R/r_{1/2,CO}<2}$)---with the highest CO line ratios---contain gas at $n_{\rm H} > 10^3$ cm$^{-3}$, which is barely present in the galaxy outskirts. The increase in $r_{31}$ with redshift, particularly in the galaxy centers, appears to result from the probability distribution function (PDF) of $n_{\rm H}$ shifting toward higher densities (see Fig.~\ref{fig:nH_hist}). Our simulated $z=2.5-3.5$ galaxies thus contain a larger fraction of dense gas in the inner regions than the $z=0.8-1.2$ galaxies, leading to higher median CO line ratios (although we note that the line ratio of individual clouds can be equally high at any redshift).
        
        We recover a much weaker trend of increasing $r_{31}$ with gas temperature, although the increase is notably steeper in the central regions, where the gas is also denser. Interestingly, the suprathermal emission (${r_{31}>1}$) is associated with the warm, dense gas at densities $\mathrm{>100\,cm^{-3}}$ and temperatures of $\sim30-100$\,K, but not the cold gas at $\sim10$ K. This population of clouds appears to stem from regions with efficient heating of the dense gas with minimal dust attenuation. Indeed, the fraction of gas with suprathermal emission is larger at higher redshifts, owing to the PDF of $\chi$ and $n_{\rm H}$ (and thereby, $\sigma_{\rm NT}$) shifting to higher values (see Figs.~\ref{fig:chi_hist}--\ref{fig:nH_hist}). Also, the CO(1--0) emission is more self-absorbed at high densities as compared to CO(3--2).

        We further investigate the impact of local star formation on the compactness of high-$J$ CO line emission and hence the steepness of the line ratio profiles. In Fig.~\ref{fig:lineratioradialprofile_sigmaSFR}, we show the CO line ratios as a function of galactocentric radius, as in Fig.~\ref{fig:lineratioradialprofile}, but now binned in SFR surface density ($\mathrm{\Sigma_{SFR}}$). We find that the profiles become steeper with increasing  $\mathrm{\Sigma_{SFR}}$, yielding more compact high-$J$ CO sizes. Therefore, at a given redshift, galaxies that are more massive and/or have larger $\mathrm{\Sigma_{SFR}}$ produce stronger impinging radiation fields (see Fig.~\ref{fig:chi_hist}), exciting CO enough to generate suprathermal emission.

        The warm, dense, and highly turbulent star-forming gas sustains increased CR ionization rates that drive suprathermal line ratios in galaxy centers (Figs.~\ref{fig:lineratioradialprofile}, \ref{fig:lineratiovsnH}, \ref{fig:chi_hist}). These findings are qualitatively consistent with previous observational and theoretical studies. Observations of local starbursts \citep{Papadopoulos_2012,Saito_2017} and z=1--3 star-forming galaxies \citep{Valentino_2020,Taylor_2025} indicate a correlation between $\Sigma_\mathrm{SFR}$ and CO excitation. This correlation results from the increase in cosmic rays and/or turbulence in combination with high gas densities \citep[e.g.][]{Meijerink_2011,Papadopoulos_2012,Bisbas_2023}. The increased collisional rates in such ISM conditions boost the CO SLEDs to higher-$J$ in PDRs \citep{Narayanan_2017,Vallini_2018}.  
        
        Several studies also support our finding that the centers of $z>1$ star-forming galaxies host the warmest and densest gas, leading to strong dependencies between galaxy compactness and CO excitation \citep[e.g.][]{Bournaud_2015,Puglisi_2021}.

        The radial decline in low-$J$ CO vs. CO(1--0) line ratios found here is consistent with resolved studies of local galaxies \citep[e.g.][]{Dumke_2001, Muraoka_2007, Muraoka_2016, Saito_2017,Leroy_2022}. For example, the CO(3--2)/CO(1--0) line ratios of several local starburst galaxies peak at well above unity in the centers, where the SFR surface densities peak \citep[e.g.][]{Dumke_2001,Muraoka_2016,Saito_2017}. Extending this to a large sample of local disk galaxies \citet{Leroy_2022} also find that both $r_{31}$ and $r_{21}$ anti-correlate with galactocentric radius and positively correlate with the local $\Sigma_\mathrm{SFR}$ and specific SFR. Even within these more typical disks, the CO(3--2)/CO(1--0) line ratio often exceeds unity in the galaxy centers, lending credibility to the high values we recover for much more highly star-forming galaxies.

        At cosmic noon there are far fewer published examples of resolved high-$J$ CO brightness profiles, especially relative to the ground transition (i.e., line ratios). The lensed, $z=2.24$ main-sequence galaxy studied in \cite{Sharon_2019} exhibited $r_\mathrm{31}$ values varying from 0.4--1.3, consistent with the inner $\sim4$ kpc of the equivalent stellar mass ($\Mstar\sim10^{10.9}\,\Msun$) and redshift bin studied here (Fig.~\ref{fig:lineratioradialprofile}). Other examples typically lack CO(1--0) observations, yet there is some evidence (from observations of multiple $J_{upp}>3$ CO transitions) to suggest increased excitation toward star-forming clumps \citep[e.g.][]{GL_2017,ALMAP_2015,Rybak_2020}. Moreover, recent studies of the integrated CO emission from z=1--3 star-forming galaxies reveal higher CO line ratios than for similar mass galaxies in the local Universe \citep{Boogaard_2020,Harrington_2021,Frias_Castillo_2023,Taylor_2025}, with dust-rich, highly star-forming galaxies exhibiting CO SLEDs that peak around $J=4-6$ \citep{Harrington_2021} due to the prevalence of gas at high densities of $n_{\rm H} \sim 10^{3-4}$\,cm$^{-3}$. 

        Our line ratios are higher than the values predicted in the simulation-based study of \citet{Olsen_2016} for the same stellar mass and redshift bin, especially in the galaxy centers. Using their spatially varying ISRF model, they predicted the CO(3--2)/CO(1--0) ratio to decrease from $\sim$0.6 in the centers to $\sim$0.5 in the outskirts. They conclude that this moderate decrease is due to the drop in $G_0$ from $\sim$25 to $\sim$5 with galactocentric radius. Although we predict a similar trend, our $\chi$ values drop by a few orders of magnitude with radius. This difference causes a much stronger decrease from suprathermal emission in the centers to values below 0.5 in the outskirts.

        We find evidence for high central line ratios, without considering the impact of AGN. Recent modeling efforts have shown that significant increases in the X-ray flux lead to integrated CO SLEDs that peak at higher-$J$ \citep{Vallini_2019}, also resulting in significantly more compact emission for CO transitions with $J\geq4-6$ \citep{Esposito_2024}. It may be that these AGN-driven effects would further compound the impact of the high central $\Sigma_\mathrm{SFR}$ and densities found here, driving even more peaked $J_\mathrm{upp}=4-5$ emission. Despite ignoring these effects in this work, the UV radiation field strength of our sample are comparable ($\chi\sim10^{1-3}$) to that of the local AGN-hosts studied in \citep{Esposito_2024} ($G_0\sim10^{1-4}$). Moreover, our CR ionization rates can vary locally rather than being fixed to the Milky Way value as in \citet{Esposito_2024}. This variation strongly affects our CO SLEDs even without a prescription for radiation from AGN.

        We emphasize that our results on the line ratios are sensitive to the modeling of cosmic rays. An overestimation of the CR ionization rate may cause the line ratios to be overpredicted. We normalize to the Milky Way ionization rate of $10^{-17}\,\mathrm{s^{-1}}$ (see Section \ref{sub:radfields}), which is significantly uncertain \citep{Indriolo_2012,Narayanan_2017}. Furthermore, our assumed scaling of $\xi_\mathrm{CR}$ with the local $\mathrm{\Sigma_{SFR}}$ may not hold in all environments. For instance, \citet{Krumholz_2023} argued that starburst systems only have moderately enhanced $\xi_\mathrm{CR}$ compared to those in the Milky Way, implying that our work may overestimate the CR heating of the gas in galaxy centers. Given the associated uncertainties and lack of observational constraints on these effects, especially at high redshift, we defer the exploration of using different CR prescriptions to future work.

    \subsection{Tracing ``sizes'' through CO observations}

            We find significant differences between the sizes of $J_\mathrm{upp}>3$ transitions vs. CO(1--0) for individual galaxies (Fig.~\ref{fig:COhighJ_over_CO10_halfradii}). For most galaxies, the higher-$J$ transitions are more compact, tracing the sharp decline in the line luminosity ratios with radius. The difference is more pronounced for $\mathrm{r_{90}}$ vs. $\mathrm{r_{1/2}}$ as the former encompasses more of the outskirts where the molecular gas is more diffuse. However, we also find a huge range in this size ratio (see percentiles in Table~\ref{tab:stats}), reflecting the diversity of the underlying galaxy morphologies and $\Sigma_\mathrm{SFR}$ which can drive strongly peaked line ratio profiles. These results imply that without additional information, different CO transitions cannot be used interchangeably in individual galaxies.

            Despite the fact that many galaxies show more compact high-$J$ emission, there is no clear statistical difference between the half-light radii of different CO transitions when averaged over the sample, in bins of stellar mass (shown in Fig.~\ref{fig:COhalfradii}). This result is driven by several effects: 1) the compactness of the sources in general, which means absolute differences in $\mathrm{r_{1/2}}$ are small, 2) the large range in the relative half-light ratios, and 3) the numerous low $\Sigma_{\rm SFR}$ galaxies with flat line ratio profiles (see Fig.~\ref{fig:lineratioradialprofile_sigmaSFR}), which average out the trend of the ones with high $\Sigma_{\rm SFR}$. We note that further complications could also affect this size comparison. Our definition of $\mathrm{r_{1/2,CO}}$ includes the entire CO surface brightness profile beyond what may be observable, potentially inflating size estimates.  Moreover, we do not include the effects of AGN, which may lead to even more peaked high-$J$ emission \citep[e.g.][]{Esposito_2024}.
            
            Given the predicted luminosity of various CO lines, we can estimate the observation time required to detect star-forming galaxies in this epoch. For example, a typical \SIMBA\ galaxy modeled at $z=2$, with a stellar mass of $10^{10-11}\,\mathrm{M_\odot}$, has integrated surface brightness ($S_{\rm CO}$) values at the CO half-light radii (for $J=1\rightarrow0$ to $J=5\rightarrow4$) of 5, 9, 11, 15, and 21 $\mathrm{mJy\,km/s}$, respectively. These predicted fluxes match stacked averages of star-forming galaxies at $z=2.0-2.7$ \citep{Boogaard_2020}. The median $S_{\rm CO}$ of $\mathrm{M_*>10^{10}\,M_\odot}$ galaxies within the respective half-light radii peaks at $J=5$ (or at even higher-$J$), similar to dust-rich and highly star-forming galaxies observed at $z\sim2$ \citep[see for e.g.][]{Harrington_2021,Frias_Castillo_2023,Taylor_2025}. For this epoch and stellar mass bin, it would thus be most efficient to measure CO half-light radii using CO(5--4). However, as shown by the surface brightness profiles in Fig.~\ref{fig:COradialprofile}, high-$J$ emission drops more steeply than CO(1--0), and this trend changes with stellar mass and redshift. We predict that tracing the $\mathrm{H_2}$ ``sizes'' of highly star-forming galaxies becomes more expensive with low-$J$ CO lines, although they trace the underlying $\mathrm{H_2}$ distribution more accurately.
    
\section{Summary}
    \label{sec:summary}

    In this work, we used the \SIMBA\ cosmological simulation \citep{Dave_2020} combined with the cloud property plus spectral line modeling pipeline, \SLICK\ \citep{Garcia_2024} to provide predictions of CO emission at $\sim$500 pc resolution in z$=$1--3 star-forming galaxies. These predictions are designed to inform the interpretation of recent and upcoming resolved observations of molecular gas. To create these predictions, we selected galaxies on and above the main sequence (and without recent merger events), extracted their gas particle (cloud) information, applied a sub-resolution model for each cloud to obtain physical properties such as the hydrogen number density ($n_{\rm H}$) and interstellar radiation field $\chi$, and used \Despotic\ to calculate CO line luminosities---from CO(1--0) to CO(5--4)---and CO plus H$_2$ abundances for each cloud by solving for equilibrium conditions. We analyzed face-on maps of the CO emission and main physical properties. Our findings can be summarized as follows. 

    One of the key questions we aimed to address was how well CO traces H$_2$ across $z$=1--3 galaxy disks, not simply in an integrated sense. Crucially, we find that the CO surface brightness does not scale uniformly with the H$_2$ mass distribution across our simulated galaxies. The CO(1--0)-to-H$_2$ conversion factor, $\alphaCO$, is $\sim$1-5 $\mathrm{M_\odot}$ (K km s$^{-1}$ pc$^2$)$^{-1}$ in the central regions of our massive simulated galaxies, but these values rise sharply beyond the central few kiloparsecs, reaching a few 10s at 5-10 kpc and several 100s beyond 10 kpc. An even sharper rise in $\alpha_\mathrm{CO(J\to J-1)}$ was found for the higher-$J$ rotational transitions. This sharp rise in conversion factors is caused mainly by the drop in CO abundance toward the galaxy outskirts but also the combined decline in the density and velocity dispersion. Thus, our findings are inconsistent with a scenario in which $\alphaCO$ is a simple function of metallicity. Instead, we find that the impacts of density, temperature, and the gas turbulence must be taken into account. Our results have important implications for observational studies as the predicted increase in $\alphaCO$ will make it extremely challenging to resolve the outer regions of the molecular ISM in $z$=1--3 galaxies (let alone the circumgalactic medium, which we have ignored here) prior to major increases to the baselines of ALMA. 

    We find that the ``sizes'' measured from CO emission are not a 1:1 reflection of the underlying molecular gas distribution. The CO half-light radii of our simulated galaxies are predicted to be 1--5 kpc, comparable to what has been measured so far for $z$=1--3 star-forming galaxies. These CO half-light radii are on average $\sim$29\% smaller than the radii enclosing half the H$_2$ mass. Based on our results, these effects could be corrected for in large observational samples but given the large range of half-light versus half-mass radii, such corrections are not necessarily applicable for individual galaxies.

    We caution that the typically used size metric of a half-light radius may miss crucial information, for instance it is not representative of the extent of the H$_2$ within these $z$=1--3 galaxy disks. Resolved observations often aim to trace emission out to twice the half-light radius, but this may still miss a significant component of the extended ISM reservoir and the circumgalactic medium. We find that on average, $\sim$30\% of the H$_2$ mass in the ISM is missed when tracing the CO emission up to twice the CO half-light radius. If only the central few kiloparsecs are of interest, this may not be an issue, as $\alphaCO$ resembles MW-like values in the centers of our (simulated) main-sequence galaxy sample. However, we predict the efficacy of CO emission in tracing $\mathrm{H_2}$ at the galaxy outskirts to be poor. With current submillimeter facilities, using CO observations to trace the full extent of the molecular gas disk would thus be unfeasible.

    Although we find only weak differences between the half-light radii measured for different CO transitions, there can be significant differences for individual galaxies, depending on their morphology and compactness of star formation. For instance, the presence of a dense and starbursting nuclear region may result in more compact mid- vs. low-$J$ emission, whereas a dense star-forming clump at the outskirts of a spiral arm may have the opposite effect.

    We find that the CO excitation increases significantly toward galaxy centers and with increasing redshift---supporting recent observational studies at $z$=1--3 \citep{Boogaard_2020,Harrington_2021,Frias_Castillo_2023}. The ratios of the CO line luminosities (relative to the ground transition) peak in the galaxy centers, with suprathermal $r_\mathrm{31}$ and $r_\mathrm{21}$ values especially for the highest-redshift bins. The line ratios also decline significantly beyond the central 1--2 kpc, reaching factors of $\sim$5-10 lower values in the outskirts. They are strongly influenced by our assumptions on the cosmic ray ionization rate modeling, as it efficiently excites CO in dense gas. For our simulated galaxies, the increase in line ratios with redshift, and toward the galaxy centers, is driven mostly by higher densities. Suprathermal values are reached in the warm, dense gas, with $T_g = 30-100$ K, and $n_{\rm H}=10^2-10^4\,\mathrm{cm^{-3}}$ as found in regions of high interstellar radiation field and cosmic ray strength.

    Our results highlight the effect of systematic uncertainties in resolving molecular gas at $z=1-3$. These uncertainties arise from an increase in the CO-to-H$_2$ conversion factor ($\alphaCO$) and decrease in the CO line ratios as a function of galactocentric radii. We find that rather than the gas-phase metallicity, cloud properties such as the hydrogen number density ($n_{\rm H}$) and the UV radiation field ($\chi$) are more directly responsible for the decorrelation of CO emission \& H$_2$ gas.\\

    \begin{acknowledgements}
    The authors thank the anonymous referee for their constructive feedback and insights on the literature. This work was conducted by RA as part of his MS Thesis at IISER Tirupati in the year 2022--2023. We would like to thank Mark Krumholz for creating the \Despotic\ Python package and for useful insights on the code. We are grateful to Anshu Gupta for valuable suggestions at the early stages of this work. RA also thanks Hollis Akins for their \SIMBA\ main-sequence code \& Sidney Lower for her galaxy projection codes. The authors acknowledge the University of Florida Research Computing for providing computational resources (specifically, the HiPerGator cluster) and support that have contributed to the research results reported in this publication. DN and KG were funded by NSF AST1909153.
    \end{acknowledgements}

\bibliographystyle{aa}
\bibliography{main.bib}

\newpage

\begin{appendix}

\onecolumn

\section{Physical properties driving changes in line excitation}

\begin{figure}[h!]
        \centering
        \includegraphics[width=0.9\textwidth]{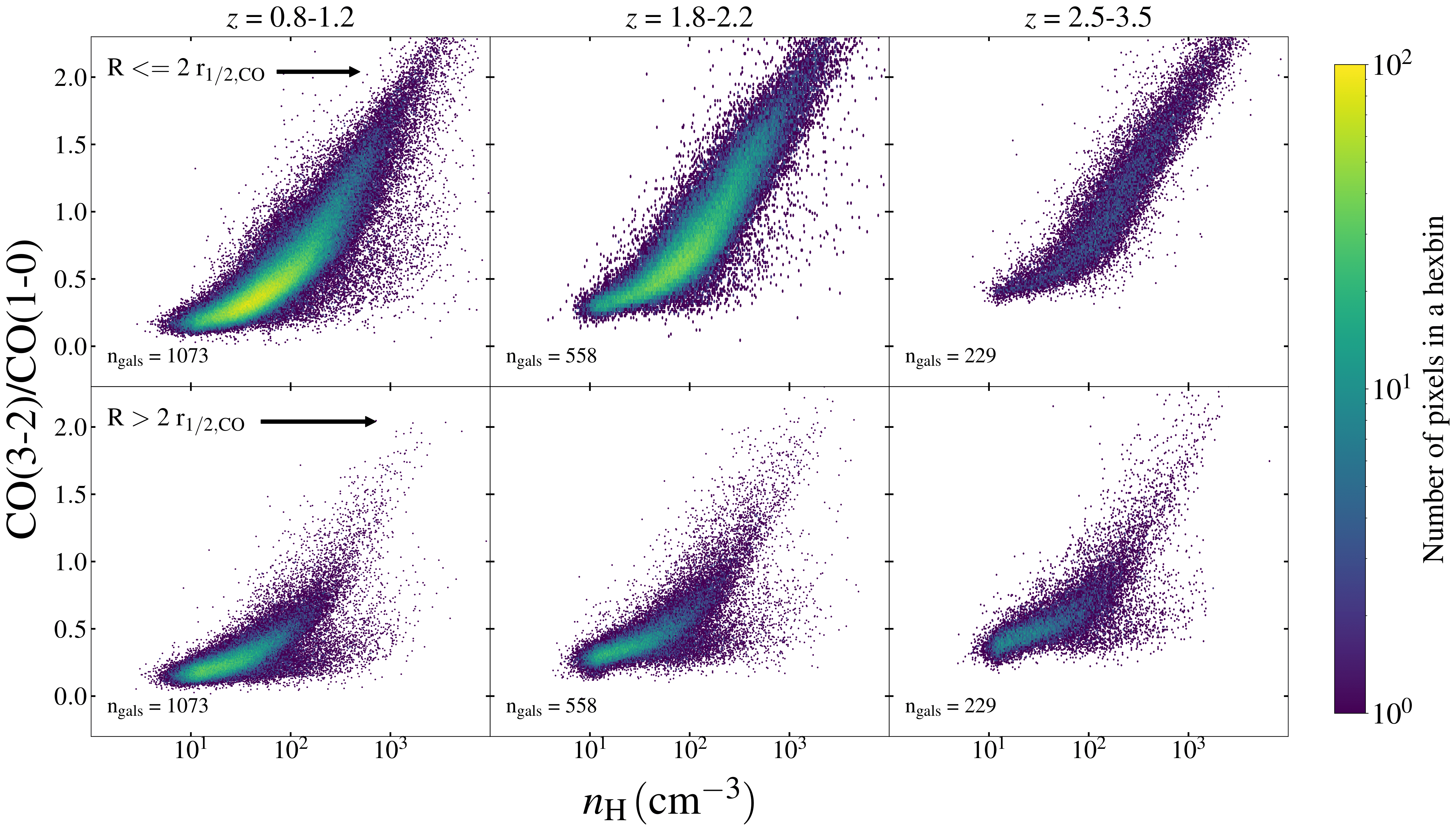}
        \caption{$r_{32}$ vs. hydrogen number density per pixel, including the pixels of all synthetic CO(1--0) and CO(3--2) maps for the entire $z$=1--3 sample. The column and row separation, and color-coding are the same as for Fig.~\ref{fig:alphaCOvsZ}. \label{fig:lineratiovsnH}}
    \end{figure}

\begin{figure}[h!]
        \centering
        \includegraphics[width=0.9\textwidth]{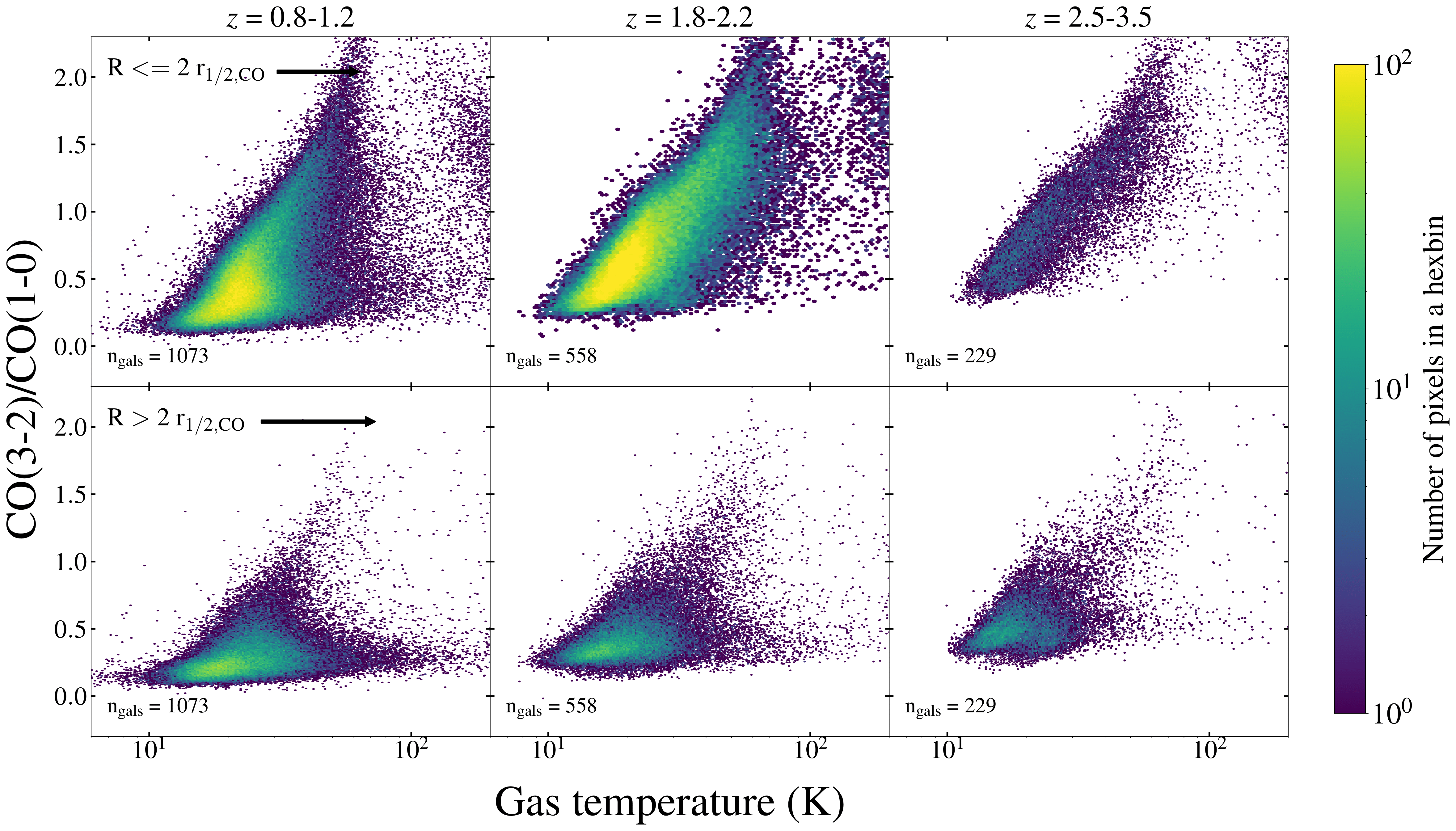}
        \caption{$r_{32}$ vs. gas temperature per pixel. The column and row separation, and color-coding are the same as for Fig.~\ref{fig:alphaCOvsZ}. \label{fig:lineratiovsTg}}
    \end{figure}

In this section, we show the dependence of the CO(3--2)/CO(1--0) line ratio on cloud properties such as the gas hydrogen number density and the temperature in Figs.~\ref{fig:lineratiovsnH}\&\ref{fig:lineratiovsTg}. Similar to Figs.~\ref{fig:alphaCOvsZ}-\ref{fig:alphaCOvssigmaNT}, we separate the pixels from our synthetic maps based on their radial distance from the galaxy - within and beyond $2\times \mathrm{r_{1/2,CO}}$. We note that the trends with $n_{\rm H}$ \& $T_{\rm g}$ described earlier are similar for different CO line ratios with the effect being more pronounced for high-$J$ transitions.

\FloatBarrier

\FloatBarrier

\begin{figure*}[h!]
        \centering
        \includegraphics[width=\textwidth]{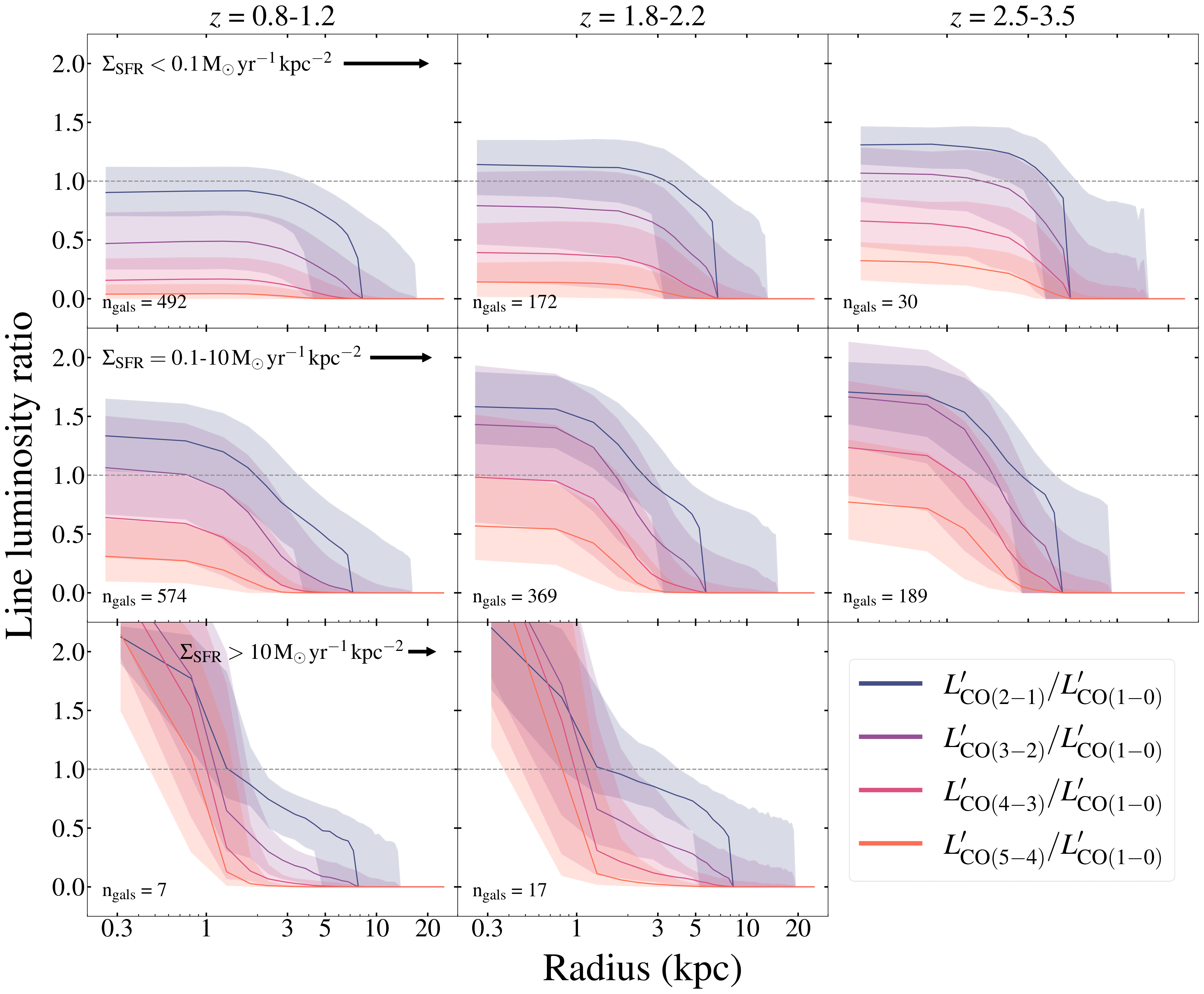}
        \caption{Average radial profile of the CO line luminosity ratios, in the same bins of redshift (columns) as Fig.~\ref{fig:lineratioradialprofile}, but the rows now in bins of $\mathrm{\Sigma_{SFR}}$. The solid lines represent the median line ratios, averaged across the entire galaxy sample shown in each panel, while the spread encompasses the 16th to 84th percentile values.\label{fig:lineratioradialprofile_sigmaSFR}}
    \end{figure*}

We also show a variation of Fig.~\ref{fig:lineratioradialprofile}, now separated in bins of different galaxy-scale star formation rate surface densities ($\mathrm{\Sigma_{SFR}=SFR/\pi r_{1/2,CO(3-2)}^2}$), in Fig.~\ref{fig:lineratioradialprofile_sigmaSFR}. We find that the line ratio profiles are steeper with increasing $\mathrm{\Sigma_{SFR}}$, as the increased star formation activity produces stronger ISRF strengths and CRs, leading to increased CO excitation. The suprathermal emission is thus mostly limited to the higher $\mathrm{\Sigma_{SFR}}$ values.

\FloatBarrier

\section{Distribution of cloud physical properties}

\begin{figure}
        \centering
        \includegraphics[width=0.7\textwidth]{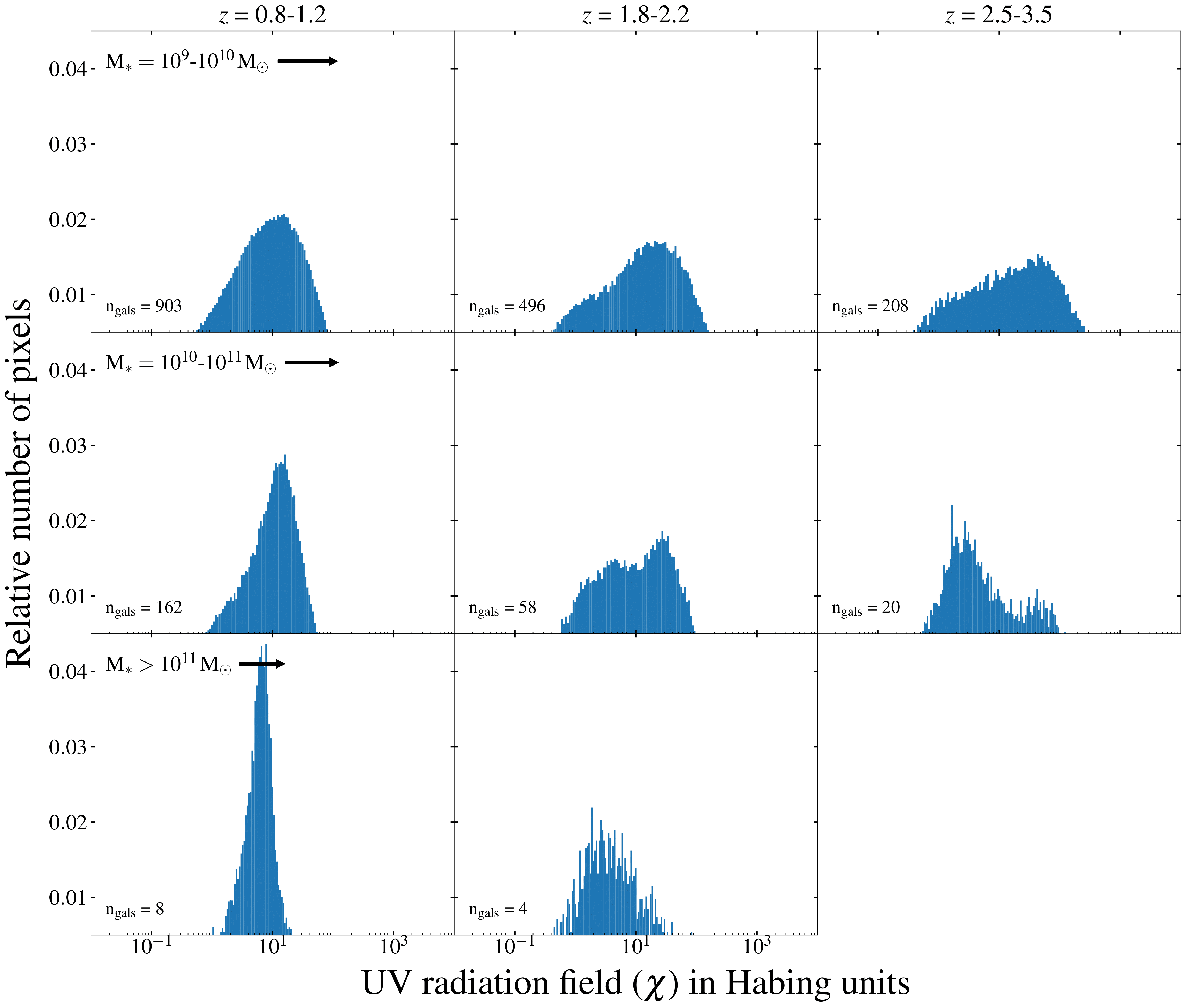}
        \caption{Probability distribution function of the interstellar radiation field ($\chi$) of each pixel in our sample of galaxies at $z$=1--3. The columns and rows indicate the same redshift, and stellar mass bins as in Fig.~\ref{fig:COradialprofile}.} \label{fig:chi_hist}
    \end{figure}

\begin{figure}
        \centering
        \includegraphics[width=0.7\textwidth]{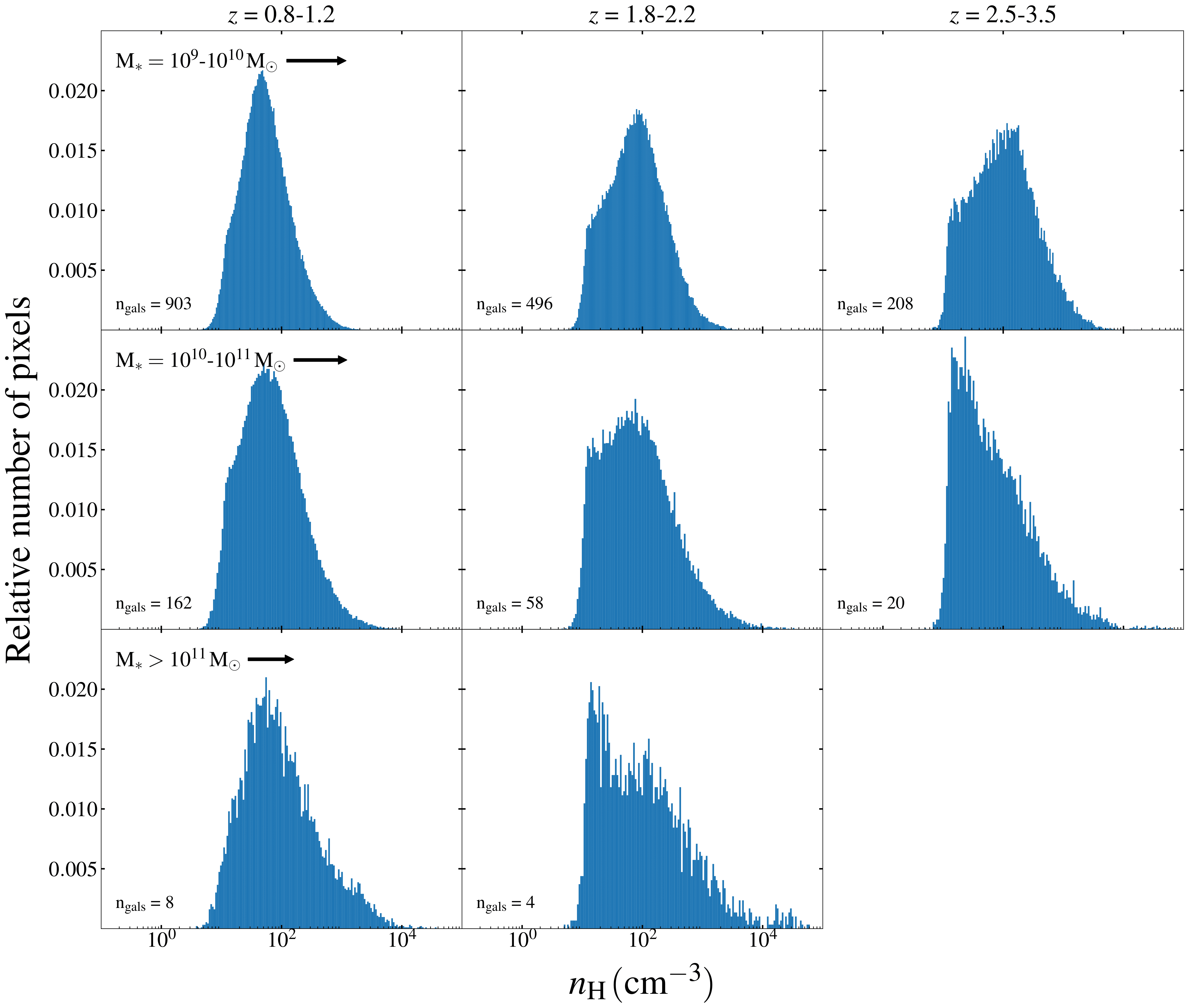}
        \caption{Probability distribution function of the hydrogen number density ($n_{\rm H}$) of each pixel in our sample of galaxies at $z$=1--3. The columns and rows indicate the same redshift, and stellar mass bins as in Fig.~\ref{fig:COradialprofile}.} \label{fig:nH_hist}
    \end{figure}

In this section, we show the probability distribution functions (PDFs) of the UV interstellar radiation field, $\chi$, and the hydrogen number density, $n_{\rm H}$, using the pixels in our synthetic maps. We bin the pixels in three stellar mass \& redshift bins each as in Fig.~\ref{fig:COradialprofile}. We then obtain the PDFs by making normalized histograms of the pixel properties ($n_{\rm H}$ \& $T_{\rm g}$). Thus, the y axis in Figs. \ref{fig:chi_hist} \& \ref{fig:nH_hist} represent the probability of pixels having a certain value of the cloud property shown on the x axis.

\end{appendix}

\end{document}